\documentclass[manuscript,two column]{aastex701}
\usepackage{amsmath}
\usepackage{graphicx}
\usepackage{natbib}
\usepackage{hyperref}

\usepackage[normalem]{ulem}

\usepackage{float}
\usepackage{physics}

\begin{document}

\title{Impact of the Initial Lorentz Factor on Kinetic Instabilities, Magnetic-Field Evolution, and Particle Acceleration in Relativistic Jets}

\author[0009-0002-6873-1259]{Ram Chandra Pageni}
\email{ramchandra.pagen@bulldogs.aamu.edu}
\affiliation{Department of Physics, Chemistry, and Mathematics, Alabama A\&M University, Normal, AL 35762, USA}

\author[0000-0001-6031-7040]{Kenichi Nishikawa}
\email{kenichi.nishikawa@aamu.edu}
\affiliation{Department of Physics, Chemistry, and Mathematics, Alabama A\&M University, Normal, AL 35762, USA}
\author[0000-0002-3064-5307]{Athina Meli}
\email{ameli@ncat.edu}
\affiliation{Physics Department, College of Science and Technology, North Carolina A\&T State University, Greensboro, NC 27411, USA}

\author[0000-0003-3932-5205]{Ioana Du\c{t}an}
\email{ioana.dutan@spacescience.ro}
\affiliation{Institute of Space Science -- INFLPR Subsidiary, Atomi\c{s}tilor 409, RO-077125 M\u{a}gurele, Romania}

\author[0000-0002-3560-0044]{Jacob Trier Frederiksen}
\email{jacob.trier.frederiksen@gmail.com}
\affiliation{Niels Bohr International Academy, The Niels Bohr Institute, Blegdamsvej 17, DK-2100, Copenhagen, Denmark}

\author[0000-0002-7101-5889]{Christoph Köhn}
\email{christoph.koehn@tu-dortmund.de}
\email{koehn@space.dtu.dk}
\affiliation{Department of Physics, Technical University of Dortmund, Otto-Hahn-Stra{\ss}e 6, 44227 Dortmund, Germany}
\affiliation{Department for Space Research and Technology (DTU Space), Technical University of Denmark, {\O}rsted Plads 348, 2800 Kgs. Lyngby, Denmark}

\author[0000-0002-6684-8691]{Nicholas R. MacDonald}
\email{nmacdona@olemiss.edu}
\affiliation{The University of Mississippi, Oxford, MS, USA}

\author[0000-0002-8131-6730]{Yosuke Mizuno}
\email{mizuno@sjtu.edu.cn}
\affiliation{School of Astronomy and Space Science, Shanghai Jiao Tong University, Shanghai 200240, China}

\begin{abstract}

We investigate how the initial Lorentz factor shapes the evolution of relativistic jets---a key open question in high-energy astrophysics. Using global three-dimensional particle-in-cell (PIC) simulations with cylindrical jet geometry on a Cartesian grid, we analyze the self-consistent evolution of relativistic electron--positron and electron--ion jets from active galactic nuclei (AGN) and their interaction with the ambient medium. We implement an injection scheme that assigns space- and time-dependent velocities to injected particles and simulate jets with initial Lorentz factors of $\gamma_{0} = 10$ and 15. Jets with the higher Lorentz factor ($\gamma_{0} = 15$) exhibit stronger electron collimation, whereas those with $\gamma_{0} = 10$ show larger growth rates of kinetic instabilities during the linear stage. In both cases, a non-oscillatory electric field accelerates and decelerates charged particles throughout the linear and nonlinear stages. Upon entering the nonlinear regime, the initial toroidal magnetic field undergoes significant amplification driven by the Weibel instability (WI), the kinetic Kelvin--Helmholtz instability (kKHI), and the mushroom instability (MI), and subsequently reorganizes into complex topologies. In the late nonlinear stage, magnetic islands with randomly swirling fields indicate a highly turbulent magnetic structure. These findings demonstrate that global 3D~PIC modeling is essential for capturing the interplay among kinetic instabilities, magnetic turbulence, and particle acceleration in relativistic outflows, bridging microscopic plasma physics with macroscopic observations of AGN jets and gamma-ray bursts.
\end{abstract}


\section{Introduction} 

Relativistic jets are highly energetic, collimated outflows of ionized plasma ejected from the vicinities of compact astrophysical objects, such as black holes, and are associated with active galactic nuclei (AGN) \citep[e.g.,][]{Blandford2019, EHT2019}, gamma-ray bursts (GRBs) \citep[e.g.,][]{Piran2004,KumarZhang2015}, and pulsar systems \citep[e.g.,][]{GaenslerSlane2006}.
These outflows are characterized by bulk Lorentz factors $\gamma > 1$
and are thought to be powered by the extraction of rotational energy
from compact objects such as black holes (BHs) or neutron stars (NSs)
\citep[e.g.,][]{Blandford1977,Blandford2019}. They can extend far beyond their host galaxies while remaining highly collimated over a broad range of spatial scales. Recent millimeter-VLBI observations resolve the M87 nucleus on event-horizon scales, while its large-scale jet remains narrowly collimated over seven orders of magnitude in distance \citep{Hada2024}. As a relativistic jet propagates through the ambient plasma, the
jet--ambient interaction excites shock-associated Weibel/filamentation
modes near the jet head and velocity-shear-driven kinetic
Kelvin--Helmholtz (kKHI) and mushroom (MI) modes at the jet boundary,
leading to magnetic-field generation and amplification. Global three-dimensional particle-in-cell (PIC) studies have begun to
capture shock- and shear-driven kinetic instabilities self-consistently
within the same cylindrical jet--ambient system \citep[e.g.,][]{Nishikawa2016a,Nishikawa2020,Meli2023}. Understanding how relativistic jet morphology evolves and how it affects non-thermal particle acceleration across a range of initial Lorentz factors with different plasma compositions remains a key open question in the field.

Jets originating from black holes and injected into the interstellar medium 
contain magnetic fields that are thought to be helical \citep[e.g.,][]{Hawley2015}. Previous 
particle-in-cell (PIC) simulation studies have extensively investigated the 
evolution of cylindrical jets across both unmagnetized \citep[e.g.,][]{Nishikawa2016a} and helically 
magnetized topologies \citep[e.g.,][]{Nishikawa2016b, Alves2018}. A key open question 
in the physics of relativistic astrophysical jets concerns the microscopic 
nature of their interaction with the ambient medium. While macroscopic RMHD models capture fluid-scale jet instabilities
such as Kelvin--Helmholtz \citep[e.g.,][]{Millas2017} and current-driven kink modes
\citep[e.g.,][]{Alves2018, Alves2019, Davelaar2020, Hu2025a,Hu2025b},
resolving kinetic instabilities (Weibel, kKHI, and MI) and the
associated particle acceleration requires a PIC approach
\citep[e.g.,][]{Nishikawa2016a,Nishikawa2020,Meli2023}. Global 3D PIC simulations have been performed to investigate the microphysical 
processes of the early evolution of electron-proton and electron-positron 
relativistic jets with helical magnetic fields, demonstrating that helical 
fields inhibit standard kinetic instabilities (WI, kKHI, and MI), but 
introduce novel instabilities linked to current-driven kinks and 
recollimation shocks \citep{Nishikawa2016b}.

Recently, \citet{Meli2023} performed global particle-in-cell (PIC) simulations of relativistic jets with a moderate initial toroidal magnetic field and showed that this field can modify the development of kinetic instabilities depending on jet plasma composition, even when those instabilities strongly amplify the magnetic field. However, the systematic dependence of jet dynamics on the initial Lorentz factor ($\gamma_{0}$) was not examined. To address this limitation, we present a suite of simulations of electron--positron ($e^{\pm}$) and electron--ion ($e^{-}$ -- $i^{+}$) jets in which $\gamma_{0}$ is varied systematically while the initial toroidal field amplitude, $B_{0}$, is held fixed. Whereas \citet{Meli2023} emphasized plasma composition, the present work isolates the role of initial Lorentz factor ($\gamma_{0}$) in the overall jet evolution and in non-thermal particle acceleration at fixed $B_{0}$, allowing us to assess its influence on instability growth rates, magnetic-field dissipation, and particle-acceleration efficiency.

Using fully kinetic three-dimensional particle-in-cell (PIC) simulations, we 
investigate the microphysical processes in relativistic jets with different 
initial Lorentz factors ($\gamma_{0} = 10$ and $\gamma_{0}=15$) and an initial 
toroidal magnetic field amplitude of $B_0 = 0.5c$ ($c = 1$ in normalized code units). Specifically, we explore 
the structural evolution of electron--positron ($e^{\pm}$) and 
electron--ion ($e^{-}$ -- $i^{+}$) jets, alongside the mechanisms of magnetic 
field generation and particle acceleration. We note that such simulations 
are currently limited by the computational box size and total simulation 
duration; nevertheless, they provide sufficient detail to investigate the 
nonlinear evolution of the jets and the amplification of the magnetic field. From our simulations, we report a systematic comparison of the evolution of magnetized jets for two different values of the initial Lorentz factor ($\gamma_{0}$), encompassing the velocity distributions of electrons and positrons/ions, the number and current densities, the corresponding electric and magnetic fields, and the efficiency of non-thermal particle acceleration.

This paper is organized as follows: In Section~\ref{sec:setup}, we describe 
the simulation setup and numerical methods used to model jet injection and 
its subsequent evolution. Section~\ref{sec:results} presents the simulation 
results, illustrating the morphological evolution of the jets alongside the 
generated current densities, induced electromagnetic fields, particle 
acceleration patterns, three-dimensional evolution of the magnetic field, and 
particle energy distributions. Finally, we summarize our findings and 
conclusions in Section~\ref{sec:summary}.

\section{Simulation setup and numerical implementation} \label{sec:setup}

We perform global three-dimensional (3D) PIC simulations of relativistic jets 
using \texttt{TRISTAN-MPI} \citep[e.g.,][]{Nishikawa2009, Nishikawa2014, 
Nishikawa2016a, Nishikawa2016b}, a parallelized 
3D electromagnetic particle-in-cell code.
\texttt{TRISTAN-MPI} uses MPI domain decomposition and a first-order
current-deposition scheme \citep{Esirkepov2001, Umeda2003}, and
self-consistently solves the full set of Vlasov--Maxwell equations
\citep[e.g.,][]{Dawson1983, Birdsall1995, Hockney1988}.
It has been applied extensively to astrophysical plasmas and relativistic
jets, including Weibel-mediated shocks
\citep[e.g.,][]{Nishikawa2003}, shear-flow instabilities
\citep[e.g.,][]{Alves2012, Nishikawa2014, Alves2015},
and cylindrical jets with helical magnetic fields
\citep[e.g.,][]{Nishikawa2016a, Nishikawa2016b}.
More recently, it has been used to investigate jets injected with an
initial toroidal magnetic field into an ambient plasma at rest
\citep{Meli2023}.

\subsection{Grid Configuration and Parallelization}

Simulations are performed on a Cartesian grid consisting of 
$1285 \times 797 \times 797$ cells. Parallelization is implemented through the 
Message Passing Interface (MPI), utilizing $N_{\rm proc} = 4356$ processors 
organized into a subdomain $1 \times 66 \times 66$ along the $x$, $y$, and $z$ directions.

We utilize approximately 4.5 million macroparticles per MPI domain. In the jet injection region, each grid cell contains 40 macroparticles, comprising 24 ambient plasma particles and 16 jet particles. This particle loading provides adequate sampling of the local velocity distributions and minimizes the numerical heating. This high particle density is critical for 
suppressing discrete particle noise and artificial collisionality, thereby 
accurately resolving the growth and saturation of kinetic instabilities in the 
collisionless regime.

The relativistic plasma jets are injected from the center of the $y$-$z$ plane 
at $x/\Delta = 100$ into the ambient plasma, initially propagating along the 
$x$-axis. The $x$-direction is managed as a single global unit to optimize the 
longitudinal calculations, while the transverse $y$-$z$ plane is heavily partitioned into 66 subdomains in each direction. We enforce absorbing
boundary conditions in the $x$-direction, while the transverse boundaries are periodic. To ensure numerical 
stability and minimize boundary artifacts, the transverse walls are located far 
enough from the jet such that the major phenomena are contained within the 
simulation domain throughout the entire simulation.

\subsection{Plasma Parameters and Jet Injection}
\begin{figure}[h]
\includegraphics[width=0.7\linewidth]{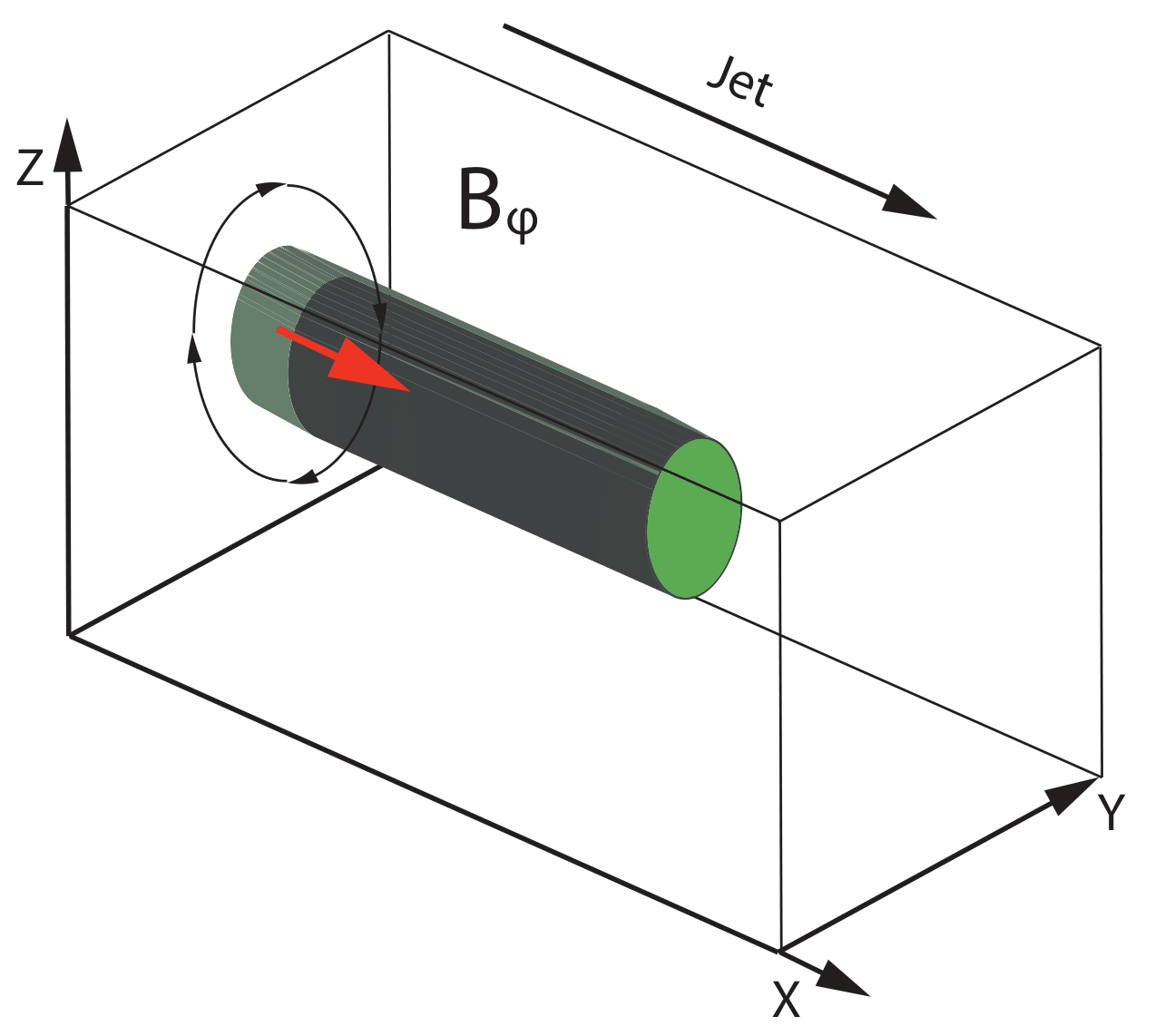}
\caption{A schematic diagram of the 3D simulation box filled with ambient plasma in which a jet is injected from the center of the $y$-$z$ plane, as indicated by the shaded cylinder \citep{Meli2023}. It is noted that the current generated in the jet is opposite to the jet propagation direction (red arrow).}
\label{fig:globaljetTb}
\end{figure}

We consider two simulation cases in which the jet and ambient plasma share the same particle composition: an electron-positron ($e^{\pm}$) plasma and an electron-ion ($e^{-}$ -- $i^{+}$) plasma. The simulation tracks a total of 19.7 billion particles, comprising 19.2 billion ambient particles and 0.5 billion jet particles. To systematically analyze the impact of the initial Lorentz factor and plasma composition, we perform a suite of four distinct simulation runs. In these simulations, we inject $e^{\pm}$ and $e^{-}$ -- $i^{+}$ jets with initial Lorentz factors of $\gamma_{0} = 10$ and $\gamma_{0} = 15$, centered at $(y_{jc}, z_{jc}) = (381, 381)$ with radii $r_{jt} = 100\Delta$, where $\Delta$ denotes the grid spacing. The injection region has a thickness of 2$\Delta$. The initial number densities measured in the jet frame are $n_{jt} = 8$ and $n_{\rm amb} = 12$ for the jet and ambient plasma, respectively. The Debye length for the ambient electrons is $\lambda_D = 0.5\Delta$, and the electron skin depth is $\lambda_{se} = c/\omega_{pe} = 10.0\Delta$, with electron plasma frequency $\omega_{pe}=(e^{2}n_{\rm amb}/\varepsilon_0 m_e)^{1/2}$. We defined each on SI unit and provide conversion between code and dimensional unit with $\epsilon_0$ = 1, 
	$\mu_0=1$, and c = $\Delta$ = 1. With this normalization, the simulation advances in discrete numerical
		iterations (\texttt{nstep}). Each iteration corresponds to a physical
		time step $\delta t=0.1\,\omega_{pe}^{-1}$. Therefore, relativistic jet particles travel approximately
		$0.1\Delta$ per time step.
		The light-crossing distance from the injector at $x/\Delta=100$ to the
		downstream boundary at $x/\Delta=1285$ is $1185\Delta$, which requires
		$(1285-100)\Delta/(0.1\Delta)=11850$ time steps.
		Thus, at $t=1000\omega_{pe}^{-1}$ (\texttt{nstep}$=10000$), light emitted
		at $t=0$ has not yet reached the right boundary.
		Since the jet partiles are injected continuously at $x/\Delta=100$, the jet head
		advances at $\simeq c$ with the earliest injected particles and lies near
		$x/\Delta\simeq1100$ at $t=1000\omega_{pe}^{-1}$. The thermal speed of jet electrons is $v_{\rm jt, th, e}$ = $\sqrt{2} \times 10^{-4}\,c$ in the jet frame, while in the ambient plasma it is $v_{\rm amb, th, e} = 0.05c$. Figure~\ref{fig:globaljetTb} presents a schematic diagram of the jet injection into the ambient medium.

The injected jet is weakly magnetized with a toroidal magnetic field strength of $B_0 = 0.5c$ . The corresponding plasma magnetization is
 \begin{equation}
    \sigma = \frac{B_0^2}{\mu_0 n_e m_e \gamma_{\mathrm{0}} c^2},
\end{equation}

where $c$ is the speed of light, $m_e$ is the electron rest mass, $\gamma_{0}$ is the initial Lorentz factor of the plasma jet, and $n_e$ is the electron number density. For fixed $B_0$, this yields $\sigma = 2.60 \times 10^{-2}$ with $\gamma_{0} = 10$ and $\sigma = 1.73 \times 10^{-2}$ with $\gamma_{0} = 15$. In the $e^{-}$ -- $i^{+}$ runs, we choose an ion-to-electron mass ratio $m_i/m_e = 4$ to control the growth of MI while retaining essential ion dynamics \citep{Meli2023}.

In this study, we apply a toroidal magnetic field of the form
\begin{equation}
B_{\phi}(r) = \frac{B_0 (r/a)}{1 + (r/a)^2},
\end{equation}
such that the axial field component $B_x = 0$. In Cartesian coordinates, the magnetic field components $B_y$ and $B_z$ are given by
\begin{equation}
    B_y(y, z) = \frac{(z - z_{jc}) B_0}{a [1 + (r/a)^2]}, \quad
    B_z(y, z) = -\frac{(y - y_{jc}) B_0}{a [1 + (r/a)^2]}.
\end{equation}

This toroidal magnetic field
$(B_{\phi})$
is imposed in the injection region (jet orifice) for each time step, which can therefore be 
treated as constant in time after the initial ramping. 

Initially, there is no current, and so
\begin{equation}
    \frac{\partial \vb{E}}{\partial t}
    = \mathrm{curl}\,\vb{B}
    = \vb{J}.
    \label{eq1}
\end{equation}
Since we apply the toroidal magnetic field $B_{\phi}$ consistent with the current in the jet,
\begin{equation}
 J_x=(\nabla\times\mathbf{B})_x
	=\frac{\partial B_z}{\partial y}-\frac{\partial B_y}{\partial z}
	=-\frac{2B_0}{a[1+(r/a)^2]^2}.
    \label{eq1}
\end{equation}
In the jet frame, all physical quantities are denoted by a prime ($'$).  The imposed toroidal field requires an axial current density $ \mathbf{J}'=\nabla'\times\mathbf{B}'$ in the jet frame. Thus, $J'_x
=
e n'_e(v'_{i,x}-v'_{e,x})
=
(\nabla'\times\mathbf{B}')_x,
$ together with the vanishing net mass flux,
$m_i n'_i v'_{i,x}
+
m_e n'_e v'_{e,x}
=
0,
$ gives the injection velocities of the jet particles:
\begin{align}
v'_{e,x}
&=
\frac{2B_0}{e n_e}
\frac{m_i}{m_i+m_e}
\frac{1}{a\bigl[1+(r/a)^2\bigr]^2},
\\
v'_{i,x}
&=
- \frac{2B_0}{e n_e}
\frac{m_e}{m_i+m_e}
\frac{1}{a\bigl[1+(r/a)^2\bigr]^2}.
\end{align}

These drift velocities are modulated by radial weight and temporal ramp in the jet frame. They are Lorentz
transformed from the jet frame to the simulation frame by using relativistic bulk velocity of the jet 
($v_{\rm bjet}$):
\begin{eqnarray}
	v_{j,x}
	&=&
	\frac{v_{j,x}'+v_{\rm bjet}}
	{1+v_{\rm bjet}\frac{v_{j,x}'}{c^2}}
	\label{eq11}
\end{eqnarray}
where $j=e,i,$ and $p$. In the simulation frame, the velocities of the jet electrons are faster than 
those of the jet positrons (ions), $v_{i,x}<v_{\mathrm{bjet}}<v_{e,x},$ producing a net negative axial current $J_x$, which generates the toroidal 
magnetic field in the clockwise direction in the injection region, as shown 
in Fig.~\ref{fig:globaljetTb}. 

In the jet orifice, the motional electric-field components for each species $j$ ($j=e$, $p$, or $i$) are applied through
		$\mathbf{E}_{\rm MEF}=-\mathbf{v}_j\times\mathbf{B}_{\phi}$,
		where $\mathbf{v}_j=v_{j,x}\hat{x}$ and $v_{j,x}$ is the $x$-component of
		the jet particle velocity (see Eq.~\ref{eq11}). In our simulations, the role of
		the motional electric field is to cancel out the Lorentz force
		($\mathbf{F}_j=q_j\mathbf{v}_j\times\mathbf{B}_{\phi}$)
		experienced by each jet species $j$ moving with $v_{j,x}$ in the toroidal
		magnetic field, such that the particles propagate straight in the
		jet-orifice region. This species-dependent electric field is applied in
		sequential order and is confined locally within the jet orifice. This setup in our simulations ensures force-free conditions in the injection 
region and yields a self-consistent current with the applied toroidal magnetic 
field $B_{\phi}$.

Since the magnetic field is applied continuously, nonlinear growing blobs are observed in \citet{Nishikawa2020}. In order to avoid this, we apply the induced electric field 
$  \mathbf{E}_{\mathrm{x}} \propto \mathbf{\nabla} \times \mathbf{B}_\phi$.
The toroidal magnetic field $B_{\phi}$ reaches its peak amplitude at $r = a$, where $a = r_{jt}/2$ is the characteristic radius of the magnetic field profile. Since the current is confined to the jet interior, the toroidal magnetic field
outside the jet decays sharply with the decay factor ($ {r_{\mathrm{jt}}}/{r}$) for $ r > r_{\mathrm{jt}}$, which enables the use of a smaller simulation box. The transverse domain is sufficiently large to ensure that self-generated azimuthal magnetic fields ($B_{\phi}$) remain negligible at the periodic boundaries ($B_{\phi} \ll B_{0}$). This prevents unphysical magnetic reconnection and artificial particle acceleration at the $y$-$z$ boundaries, confirming that the observed dynamics are driven solely by jet-ambient plasma interactions.

\begin{figure*}[t!]
    \centering

\begin{minipage}{0.48\textwidth}
    \makebox[\linewidth]{\phantom{\textbf{(a)}\hspace{1.0cm}}\hfill\textbf{${e^{\pm}}$ jet}\hfill\textbf{(a)}\hspace{1.0cm}}\\[-0.06cm]
    \includegraphics[width=\textwidth]{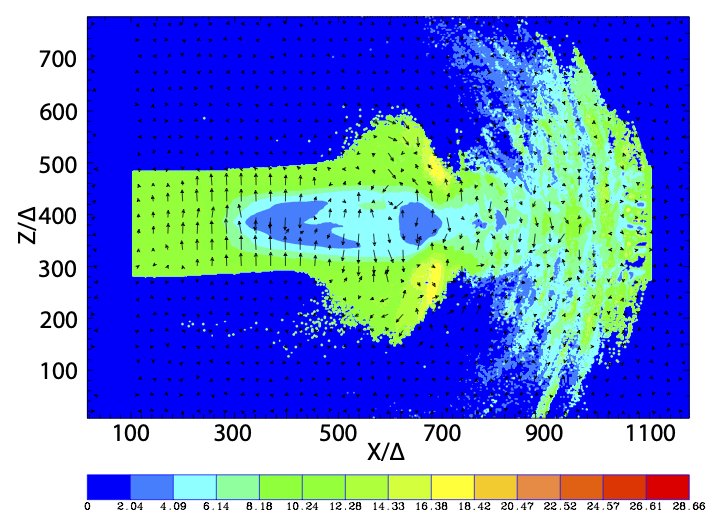}
\end{minipage}
\hfill
\begin{minipage}{0.48\textwidth}
    \makebox[\linewidth]{\phantom{\textbf{(b)}\hspace{1.0cm}}\hfill\textbf{ $e^{-}$ -- $i^{+}$ jet}\hfill\textbf{(b)}\hspace{1.0cm}}\\[-0.06cm]
    \includegraphics[width=\textwidth]{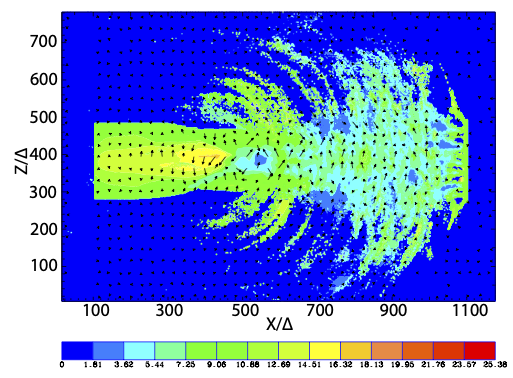}
\end{minipage}

\vspace{-0.06cm}%

\begin{minipage}{0.48\textwidth}
    \makebox[\linewidth][r]{\textbf{(c)}\hspace{1.0cm}}\\[-0.06cm]
    \includegraphics[width=\textwidth]{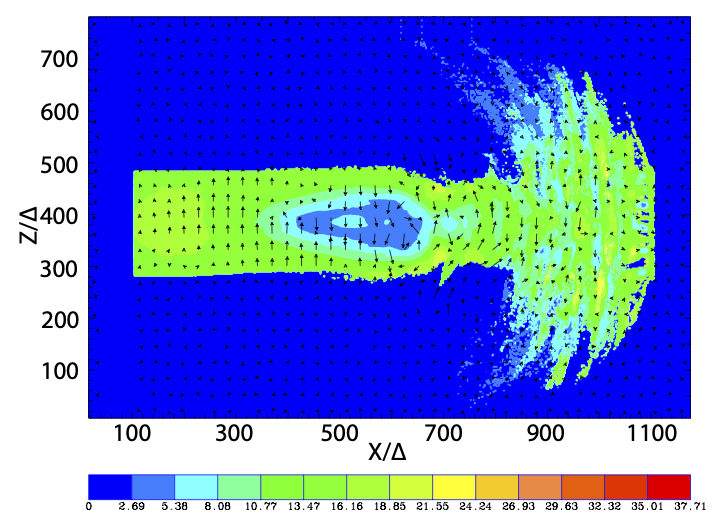}
\end{minipage}
\hfill
\begin{minipage}{0.48\textwidth}
    \makebox[\linewidth][r]{\textbf{(d)}\hspace{1.0cm}}\\[-0.08cm]
    \includegraphics[width=\textwidth]{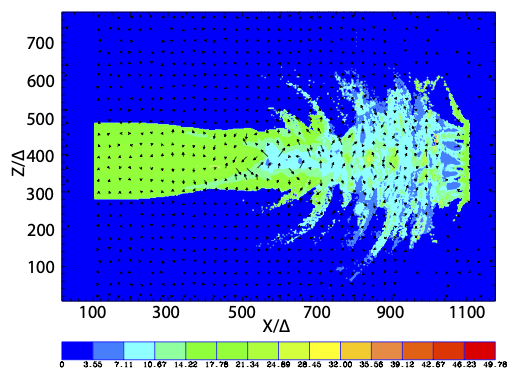}
\end{minipage}
\vspace{-0.08cm}
\caption{Lorentz factor of electrons
    at $y/\Delta = 381$ and $t = 1000\,\omega_{\mathrm{pe}}^{-1}$.
    Upper panels ($\gamma_{0} = 10$): maximum values are (a) $28.66$ and
    (b) $25.38$; bottom panels ($\gamma_{0} = 15$): (c) $37.71$ and
    (d) $49.78$.
    Black arrows indicate the in-plane magnetic field components ( $B_x$, $B_z$).}
    \label{fig:lorentz_electrons_combined}
\end{figure*}

To investigate the nonlinear stage of the jet evolution, we follow the jet for a sufficiently long duration of $t_{\max} = 1000\,\omega_{pe}^{-1}$.
We assume a top-hat density profile for both the $e^{\pm}$ and $e^{-}$--$i^{+}$ jets. In the jet frame, the injected jet number density is spatially uniform inside the jet radius and vanishes outside,

\begin{equation}
n_j(r)=
\begin{cases}
n_{jt}, & r \le r_{jt},\\
0, & r > r_{jt},
\end{cases}
\label{eq:tophat}
\end{equation}
where $r=\sqrt{(y-y_{jc})^{2}+(z-z_{jc})^{2}}$ is the cylindrical radius measured from the jet axis, $r_{jt}=100\Delta$ is the jet radius, and $n_{jt}$ is the uniform jet density. Although the shape of a real jet is far more complex, the present results represent the first step in a series of advanced numerical investigations, including the implementation of Gaussian or Lorentzian (non-top-hat) profiles that better approximate the jet density profiles.

\begin{figure*}[t!]
    \centering

    \begin{minipage}{0.48\textwidth}
        \makebox[\linewidth]{\phantom{\textbf{(a)}\hspace{1.0cm}}\hfill\textbf{$e^{\pm}$ jet}\hfill\textbf{(a)}\hspace{1.0cm}}\\[-0.06cm]
        \includegraphics[width=\textwidth]{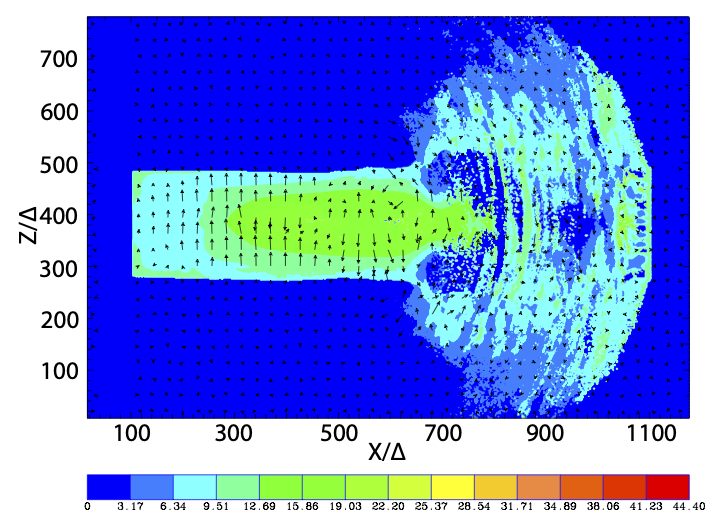}
    \end{minipage}
    \hfill
    \begin{minipage}{0.48\textwidth}
        \makebox[\linewidth]{\phantom{\textbf{(b)}\hspace{1.0cm}}\hfill\textbf{ $e^{-}$ -- $i^{+}$ jet}\hfill\textbf{(b)}\hspace{1.0cm}}\\[-0.06cm]
        \includegraphics[width=\textwidth]{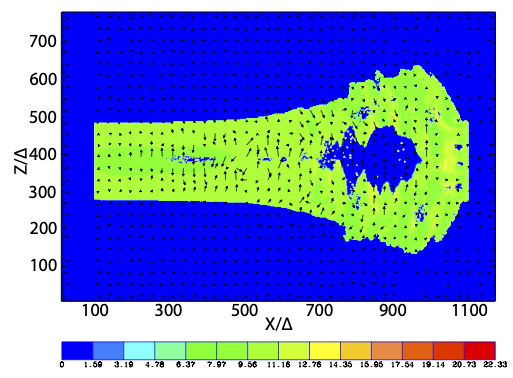}
    \end{minipage}
    \vspace{-0.06cm}

    \begin{minipage}{0.48\textwidth}
        \makebox[\linewidth][r]{\textbf{(c)}\hspace{1.0cm}}\\[-0.06cm]
        \includegraphics[width=\textwidth]{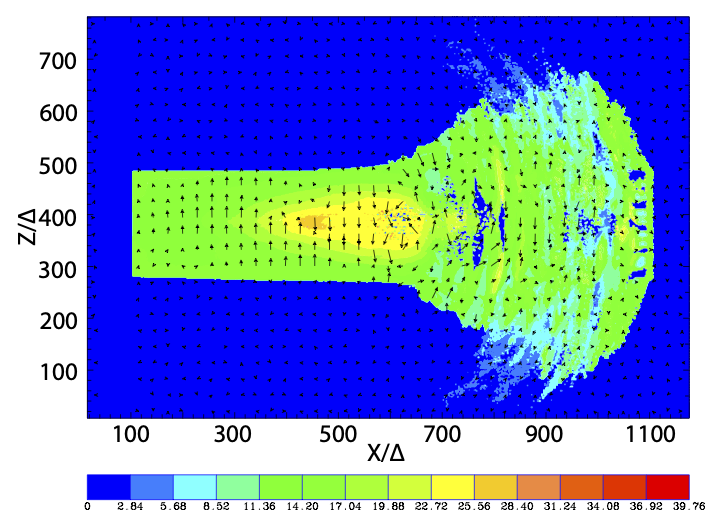}
    \end{minipage}
    \hfill
    \begin{minipage}{0.48\textwidth}
        \makebox[\linewidth][r]{\textbf{(d)}\hspace{1.0cm}}\\[-0.06cm]
        \includegraphics[width=\textwidth]{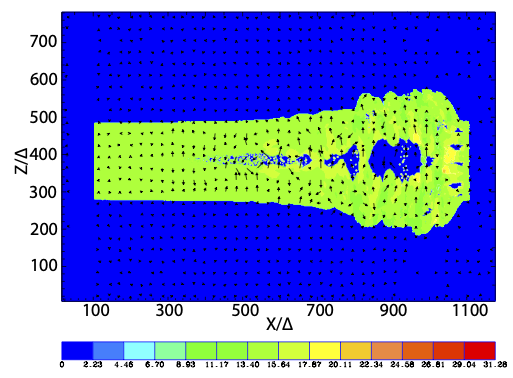}
    \end{minipage}

    \caption{Lorentz factor of positrons and ions 
    at $y/\Delta = 381$ and $t = 1000\,\omega_{\mathrm{pe}}^{-1}$.
    Upper panels ($\gamma_{0} = 10$): maximum values are (a) $44.40$ and
    (b) $22.33$; bottom panels ($\gamma_{0} = 15$): (c) $39.76$ and
    (d) $31.28$.
    Black arrows indicate the in-plane magnetic field components ( $B_x$, $B_z$).}
    \label{fig:lorentz_positrons_ions_combined}
\end{figure*}

\section{Simulation Results}
\label{sec:results}
\begin{figure*}[t!]
    \centering

    \makebox[\linewidth]{%
        \makebox[0.46\linewidth]{\phantom{\textbf{(a)}\hspace{1.0cm}}\hfill$\gamma_{0} = 10$\hfill\textbf{(a)}\hspace{1.0cm}}%
        \makebox[0.08\linewidth][c]{\textbf{$e^{\pm}$ jet}}%
        \makebox[0.42\linewidth]{\phantom{\textbf{(b)}\hspace{1.0cm}}\hfill$\gamma_{0} = 15$\hfill\textbf{(b)}\hspace{1.0cm}}%
    }\\[-0.06cm]

    \begin{minipage}{0.48\linewidth}
        \includegraphics[width=\linewidth]{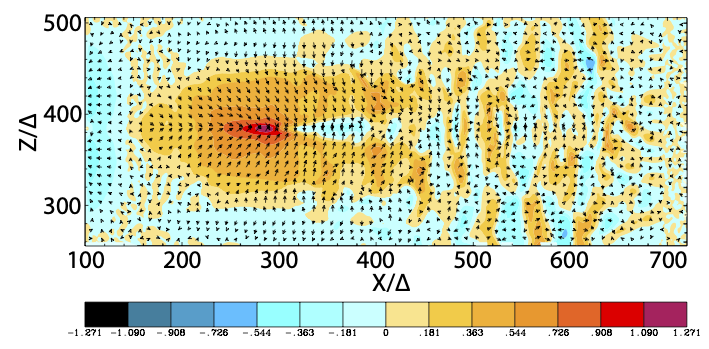}
    \end{minipage}
    \hfill
    \begin{minipage}{0.48\linewidth}
        \includegraphics[width=\linewidth]{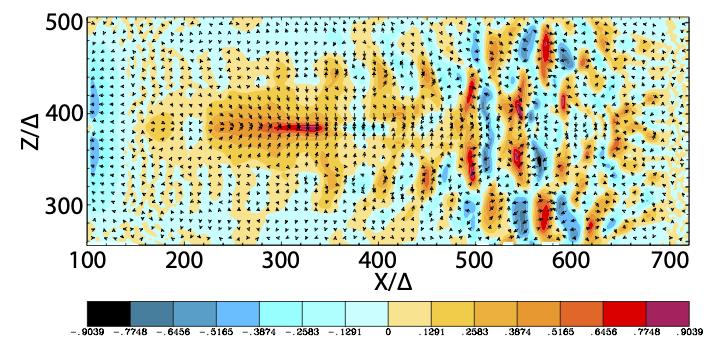}
    \end{minipage}

    \vspace{-0.06cm}

\begin{minipage}{0.48\textwidth}
  \centering
  \makebox[\linewidth][r]{\textbf{(c)}\hspace{1.0cm}}\\
  \vspace{-0.1cm}
  \includegraphics[width=\textwidth]{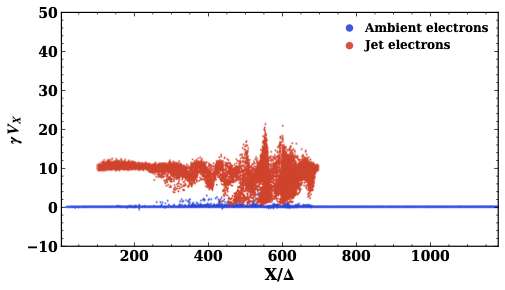}
\end{minipage}
\hfill
\begin{minipage}{0.48\textwidth}
  \centering
  \makebox[\linewidth][r]{\textbf{(d)}\hspace{1.0cm}}\\
  \vspace{-0.1cm}
  \includegraphics[width=\textwidth]{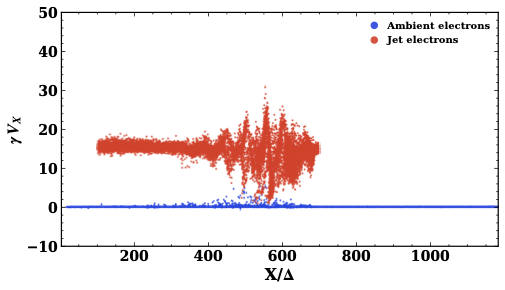}
\end{minipage}

\vspace{0.2cm}

    \makebox[\linewidth]{%
        \makebox[0.48\linewidth]{\phantom{\textbf{(e)}\hspace{1.0cm}}\hfill\hfill\textbf{(e)}\hspace{1.0cm}}%
        \makebox[0.08\linewidth][c]{\textbf{ $e^{-}$ -- $i^{+}$ jet}}%
        \makebox[0.42\linewidth]{\phantom{\textbf{(f)}\hspace{1.0cm}}\hfill\hfill\textbf{(f)}\hspace{1.0cm}}%
    }\\[-0.06cm]

    \begin{minipage}{0.48\linewidth}
        \includegraphics[width=\linewidth]{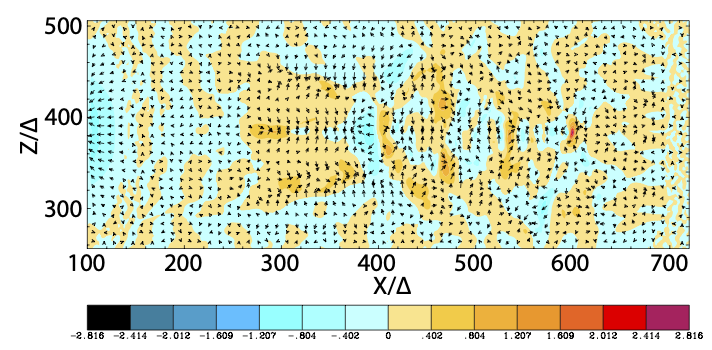}
    \end{minipage}
    \hfill
    \begin{minipage}{0.48\linewidth}
        \includegraphics[width=\linewidth]{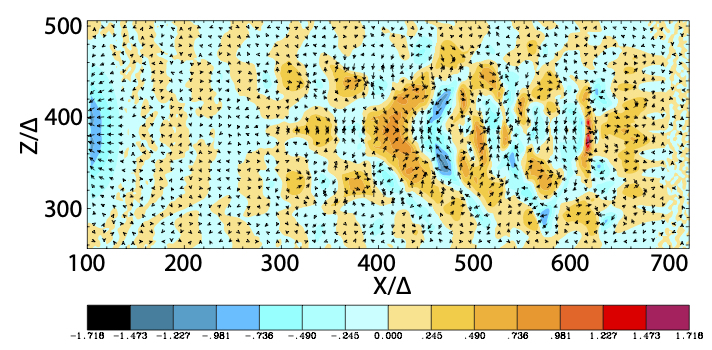}
    \end{minipage}

    \vspace{-0.06cm}

\begin{minipage}{0.48\textwidth}
  \centering
  \makebox[\linewidth][r]{\textbf{(g)}\hspace{1.0cm}}\\
  \vspace{-0.02cm}
  \includegraphics[width=\textwidth]{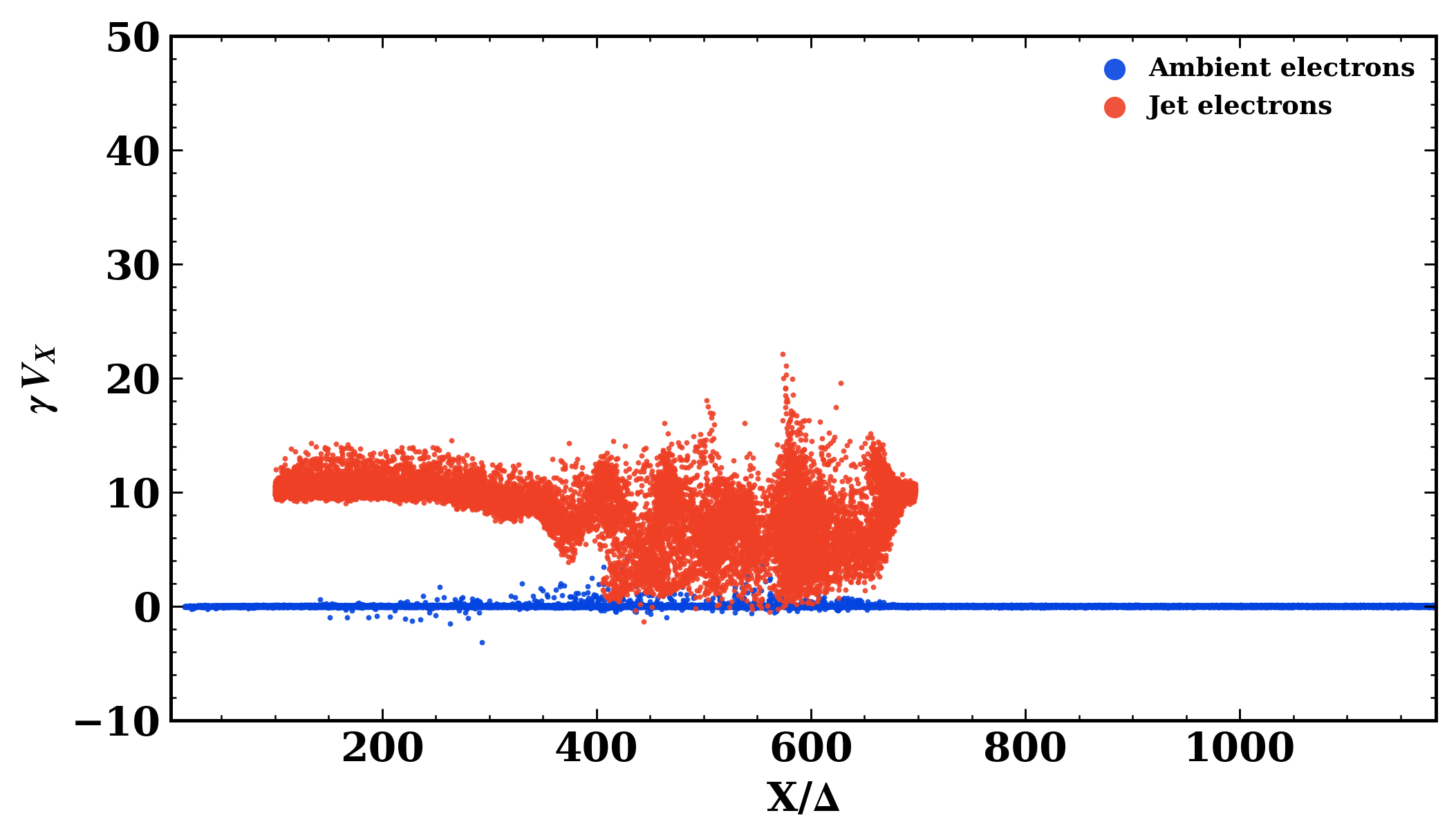}
\end{minipage}
\hfill
\begin{minipage}{0.48\textwidth}
  \centering
  \makebox[\linewidth][r]{\textbf{(h)}\hspace{1.0cm}}\\
  \vspace{-0.1cm}
  \includegraphics[width=\textwidth]{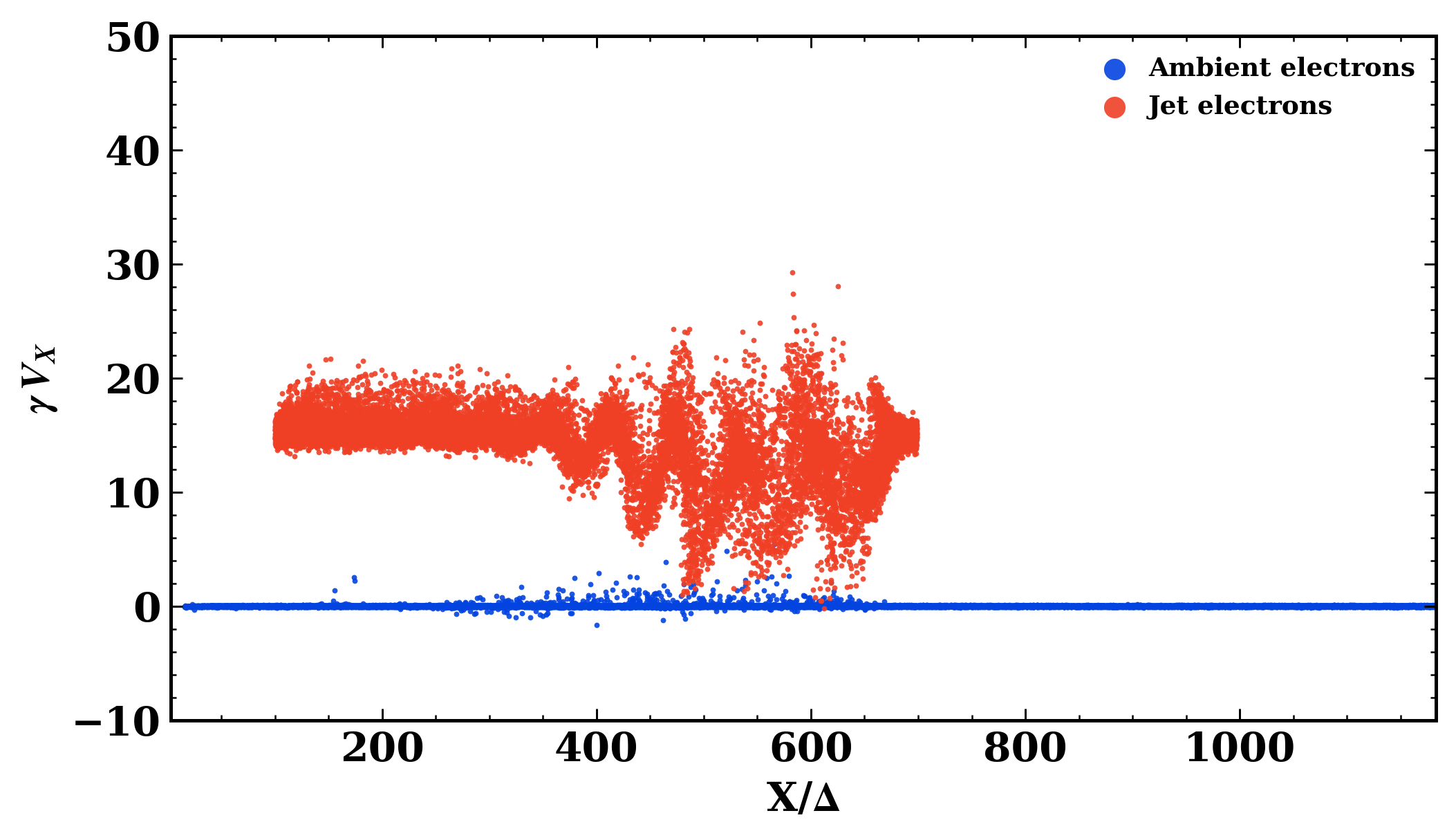} 
\end{minipage}

\vspace{0.1cm}

\caption{Electric field component $E_{\rm x}$ and the corresponding
$x$--$\gamma v_{x}$ phase-space distributions at $t = 600\,\omega_{\mathrm{pe}}^{-1}$. The maximum values of $E_{\rm x}$ are $\pm 1.271$ (a),
$\pm 0.9039$ (b), $\pm 2.816$ (e) and
$\pm 1.718$ (f). The black arrows indicate the in-plane magnetic field
components ( $B_x$, $B_z$). }
\label{fig:Ex_phase_4x2_combined}
\end{figure*}

\begin{figure*}[t!]
    \centering

    \makebox[\linewidth]{%
        \makebox[0.46\linewidth]{\phantom{\textbf{(a)}\hspace{1.0cm}}\hfill$\gamma_{0} = 10$\hfill\textbf{(a)}\hspace{1.0cm}}%
        \makebox[0.08\linewidth][c]{\textbf{$e^{\pm}$ jet}}%
        \makebox[0.42\linewidth]{\phantom{\textbf{(b)}\hspace{1.0cm}}\hfill$\gamma_{0} = 15$\hfill\textbf{(b)}\hspace{1.0cm}}%
    }\\[-0.06cm]
\begin{minipage}{0.48\textwidth}
  \centering
  \includegraphics[width=\textwidth]{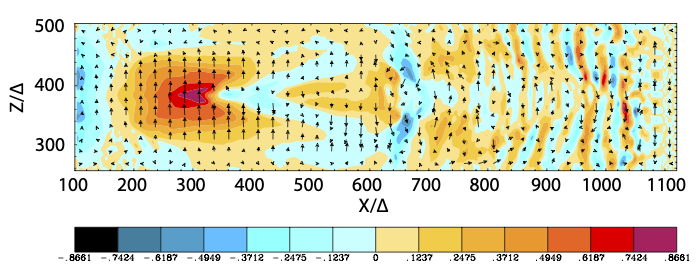}
\end{minipage}
\hfill
\begin{minipage}{0.48\textwidth}
  \centering
  \includegraphics[width=\textwidth]{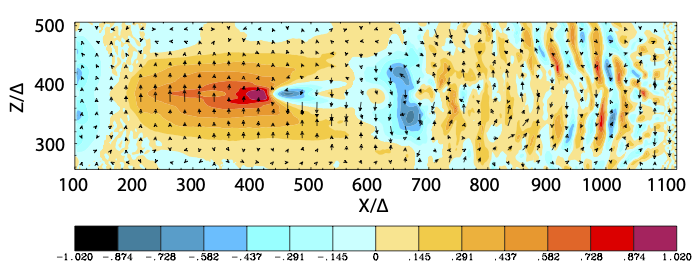}
\end{minipage}

\vspace{-0.1cm}

\begin{minipage}{0.48\textwidth}
  \centering
  \makebox[\linewidth][r]{\textbf{(c)}\hspace{1.0cm}}\\
  \vspace{-0.1cm}
  \includegraphics[width=\textwidth]{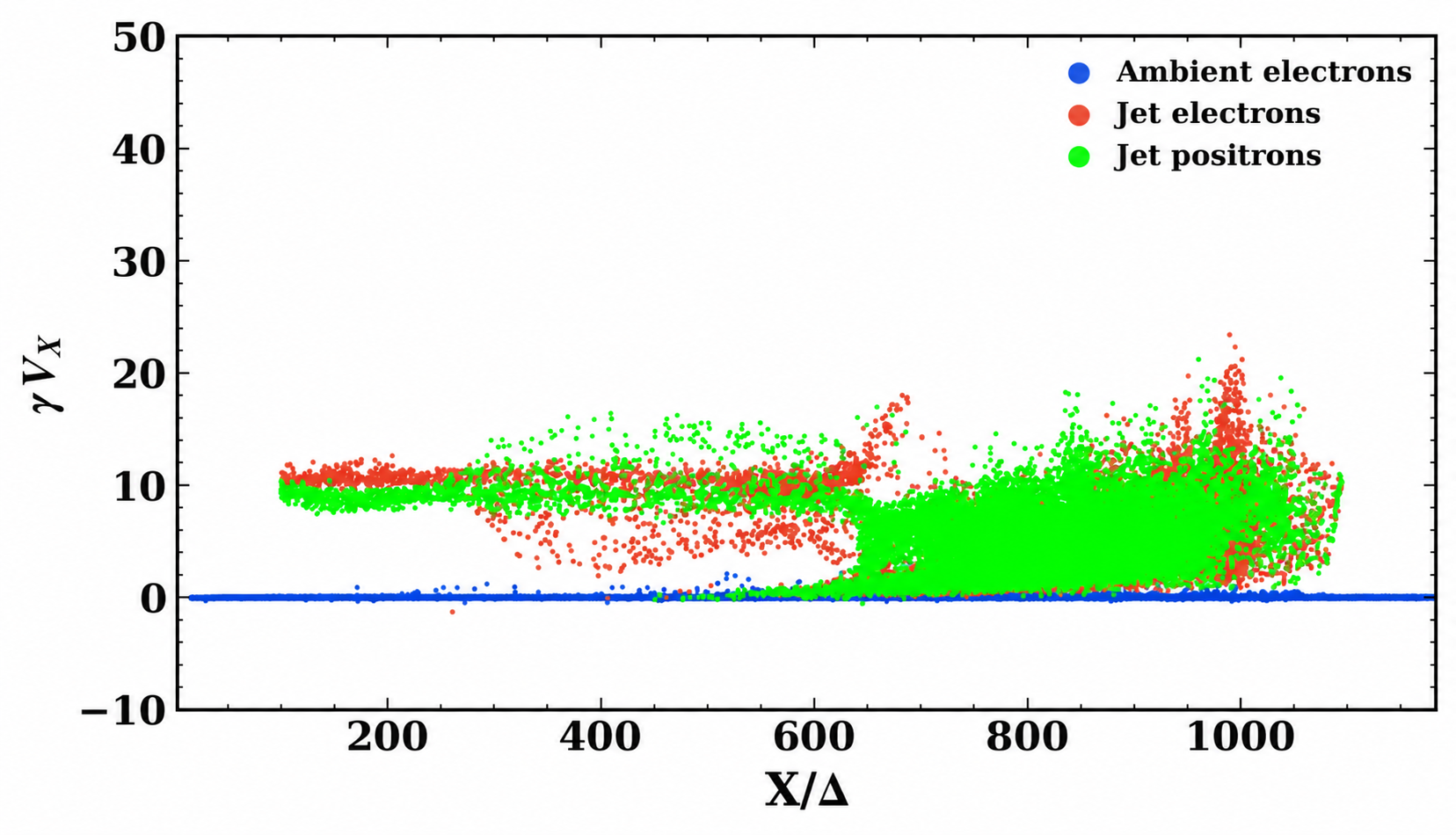}
\end{minipage}
\hfill
\begin{minipage}{0.48\textwidth}
  \centering
  \makebox[\linewidth][r]{\textbf{(d)}\hspace{1.0cm}}\\
  \vspace{-0.1cm}
  \includegraphics[width=\textwidth]{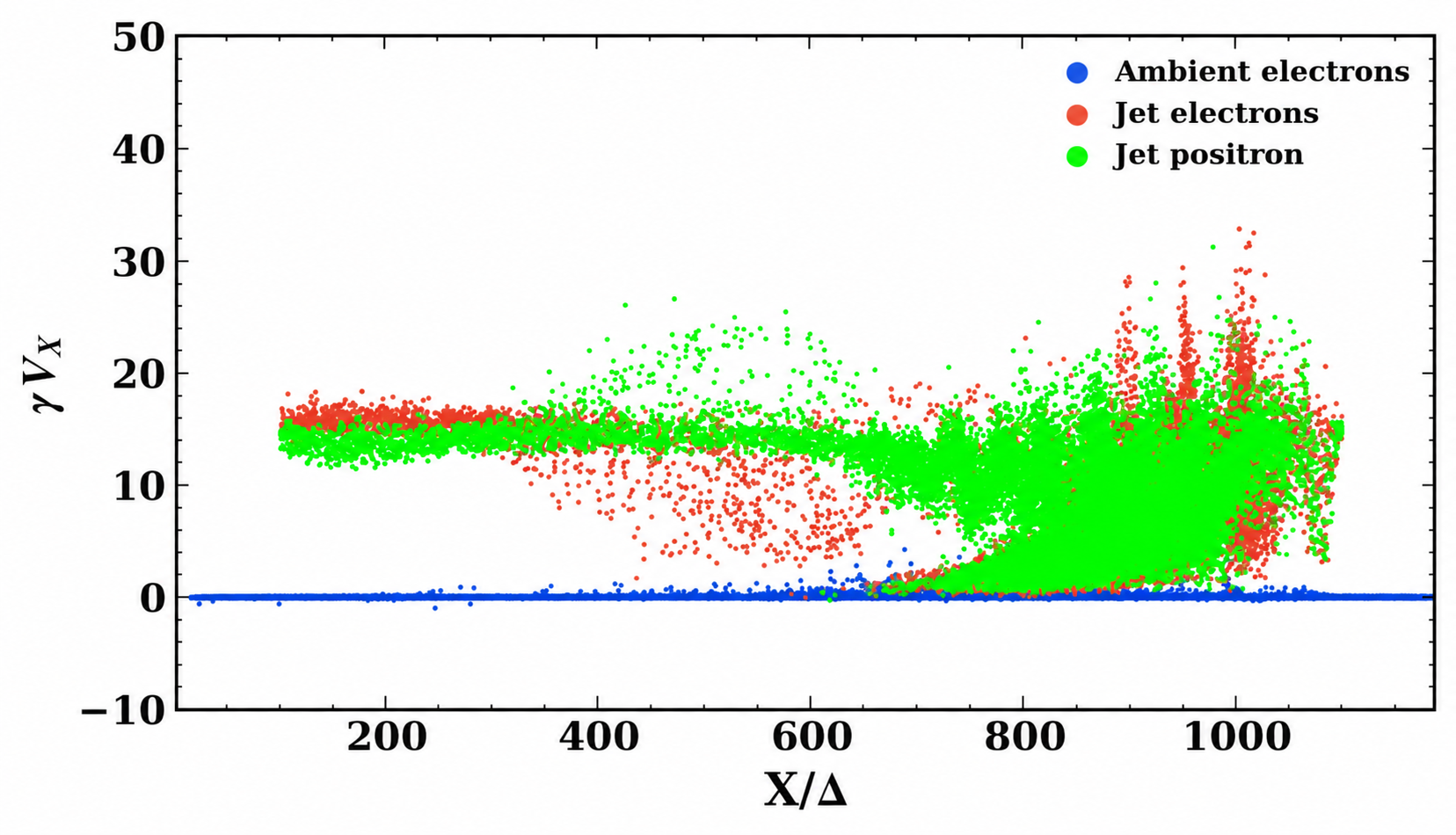}
\end{minipage}

\vspace{-0.1cm}

\caption{Electric field component $E_{\rm x}$ in the $x$--$z$ plane at $y/\Delta = 381$ for $t = 1000\,\omega_{\mathrm{pe}}^{-1}$, overlaid with
in-plane magnetic field components ( $B_x$, $B_z$), and the corresponding
$x$--$\gamma v_{x}$ phase-space distributions of electrons and positrons (c, d).
$E_{\rm x}$ profiles with extrema $\pm 0.8661$ (a) and $\pm 1.020$ (b).}
\label{fig:Ex_phasepa_combined_all}
\end{figure*}

\begin{figure*}[t!]
\centering

    \makebox[\linewidth]{%
        \makebox[0.46\linewidth]{\phantom{\textbf{(a)}\hspace{1.0cm}}\hfill$\gamma_{0} = 10$\hfill\textbf{(a)}\hspace{1.0cm}}%
        \makebox[0.08\linewidth][c]{\textbf{ $e^{-}$ -- $i^{+}$ jet}}%
        \makebox[0.42\linewidth]{\phantom{\textbf{(b)}\hspace{1.0cm}}\hfill$\gamma_{0} = 15$\hfill\textbf{(b)}\hspace{1.0cm}}%
    }\\[-0.06cm]
\begin{minipage}{0.48\textwidth}
  \centering
  \includegraphics[width=\textwidth]{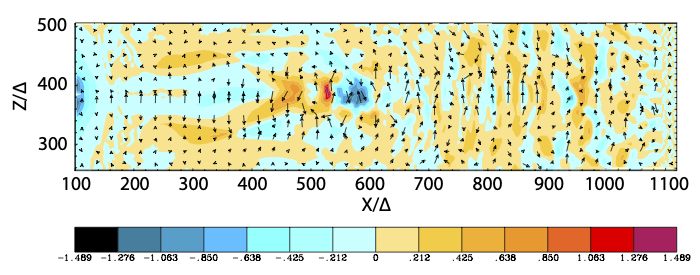}
\end{minipage}
\hfill
\begin{minipage}{0.48\textwidth}
  \includegraphics[width=\textwidth]{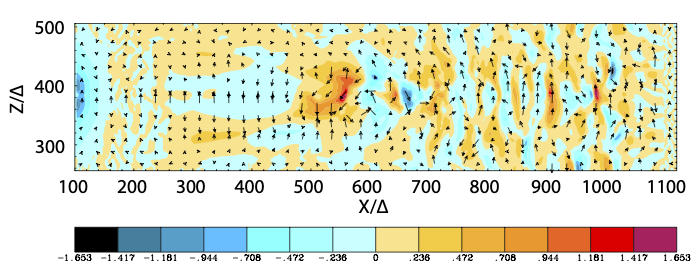}
\end{minipage}

\vspace{-0.1cm}

\begin{minipage}{0.48\textwidth}
  \centering
  \makebox[\linewidth][r]{\textbf{(c)}\hspace{1.0cm}}\\
  \vspace{-0.1cm}
 \includegraphics[width=\textwidth]{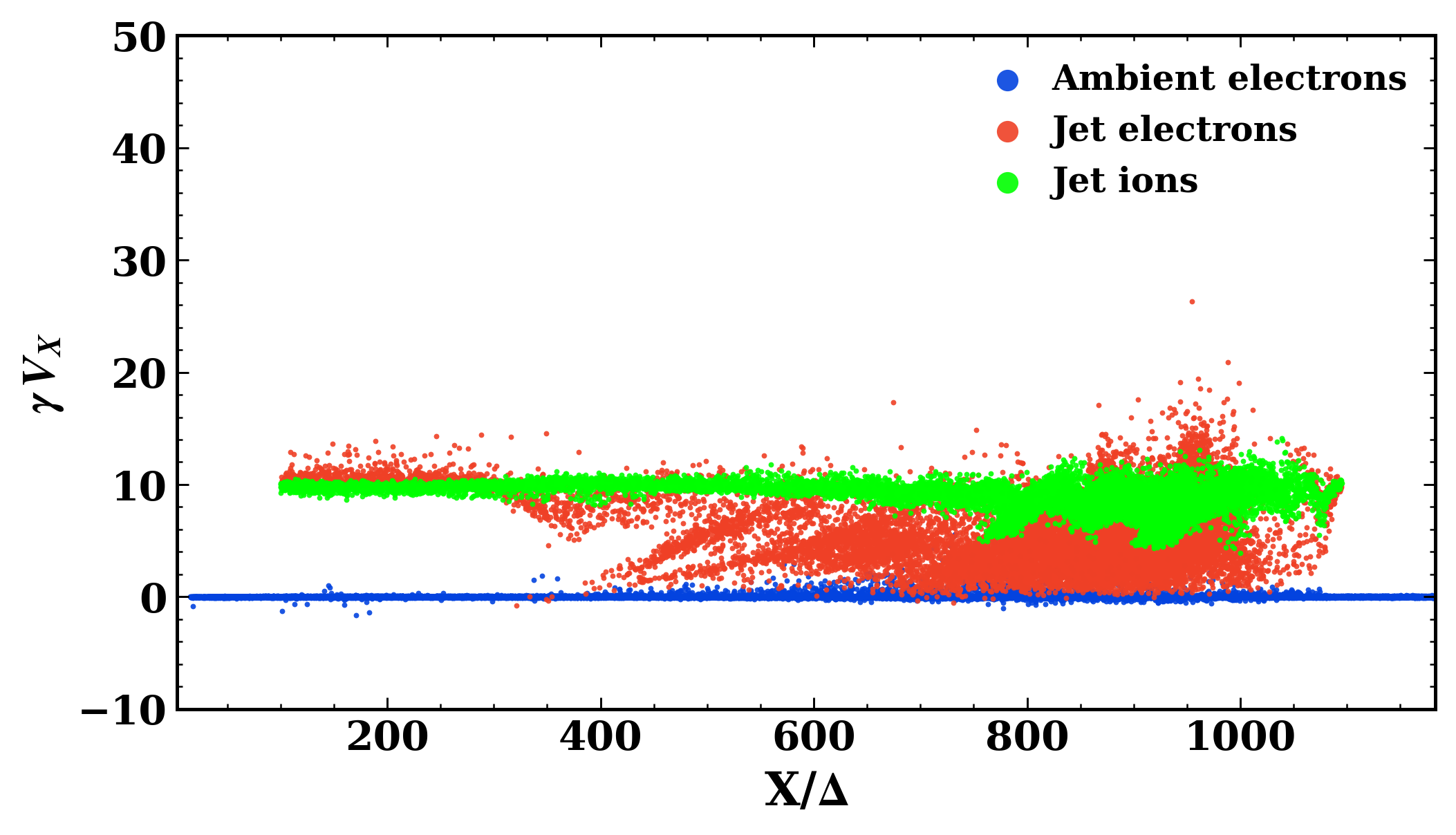}
\end{minipage}
\hfill
\begin{minipage}{0.48\textwidth}
  \centering
  \makebox[\linewidth][r]{\textbf{(d)}\hspace{1.0cm}}\\
  \vspace{-0.1cm}
  \includegraphics[width=\textwidth]{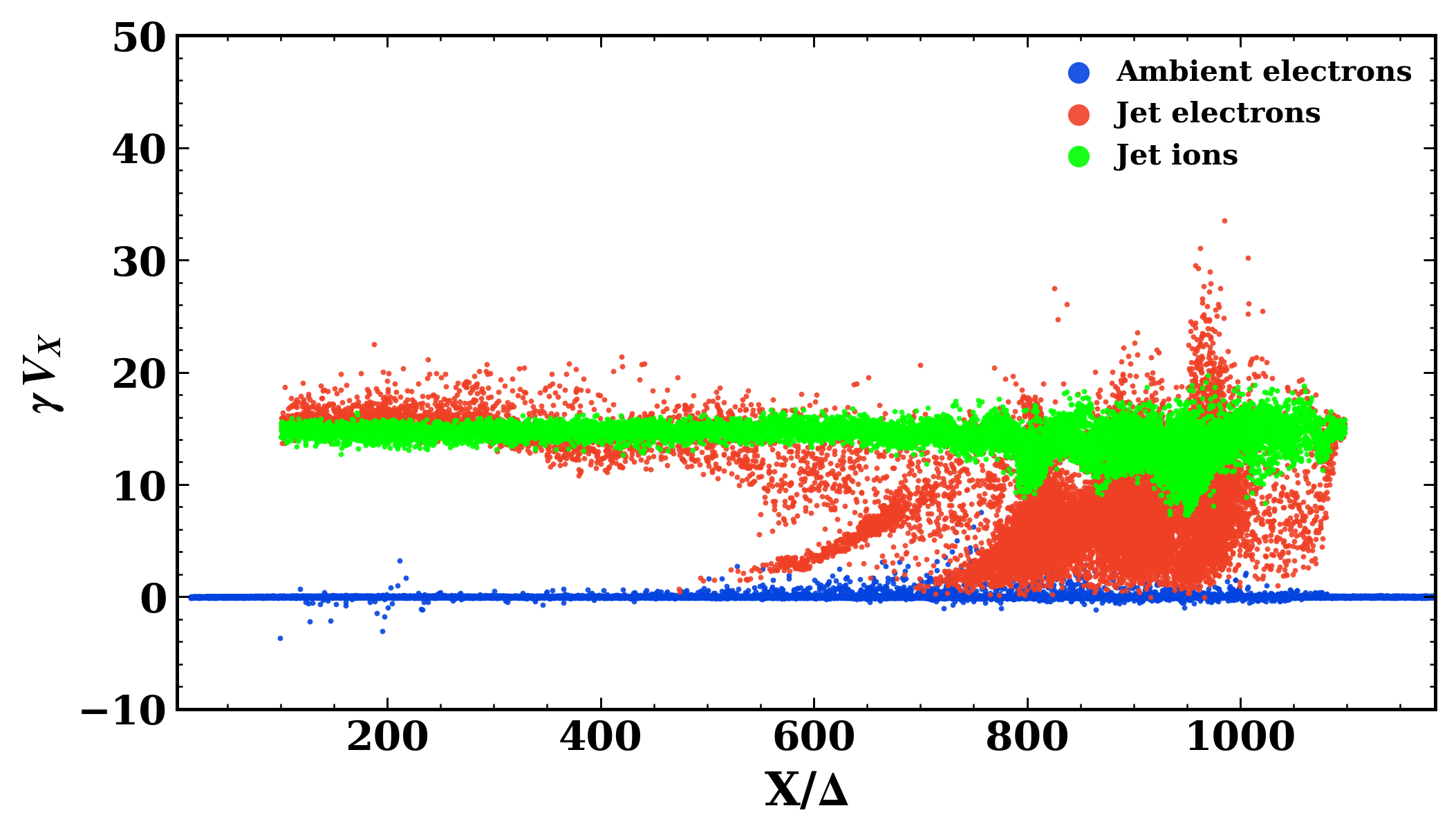}
\end{minipage}

\vspace{-0.1cm}

\caption{Electric field component $E_{\rm x}$ in the $x$--$z$ plane at $y/\Delta = 381$, overlaid with
in-plane magnetic field components ( $B_x$, $B_z$), and the corresponding
$x$--$\gamma v_{x}$ phase-space distributions of electrons and ions (c, d) at
$t = 1000\,\omega_{\mathrm{pe}}^{-1}$.
$E_{\rm x}$ profiles with extrema $\pm 1.489$ (a) and $\pm 1.653$ (b).
}
\label{fig:Ex_phas_ra_combined_comparison}
\end{figure*}

We present simulation results for weakly magnetized $e^{\pm}$ and
 $e^{-}$ -- $i^{+}$ jets propagating into an unmagnetized ambient plasma.
Our primary aim is to examine how the dynamical evolution of the jets
and their interaction with the surrounding medium vary with the initial
Lorentz factor ($\gamma_{0}$) and plasma composition.
Figures~\ref{fig:lorentz_electrons_combined}--\ref{fig:Btotyzisland_combined_2by2}
illustrate how the initial Lorentz factor and plasma composition shape
the overall jet--plasma interaction and the resulting jet morphology.
Figures~\ref{fig:Ex_phase_4x2_combined}--\ref{fig:combined_energy_dist}
highlight how these parameters influence the development of electromagnetic
field structures, the mechanisms of particle acceleration, and the resulting
particle energy distributions.

\subsection{Global Jet Structure and Kinetic Instabilities}
\label{sec:str}

\begin{figure*}[t!]
    \centering

    \makebox[\linewidth]{%
        \makebox[0.48\linewidth]{\phantom{\textbf{(a)}\hspace{1.0cm}}\hfill$\gamma_{0} = 10$\hfill\textbf{(a)}\hspace{1.0cm}}%
        \makebox[0.08\linewidth][c]{\textbf{$e^{\pm}$ jet}}%
        \makebox[0.42\linewidth]{\phantom{\textbf{(b)}\hspace{1.0cm}}\hfill$\gamma_{0} = 15$\hfill\textbf{(b)}\hspace{1.0cm}}%
    }\\[-0.06cm]

    \begin{minipage}{0.48\linewidth}
        \includegraphics[width=\linewidth]{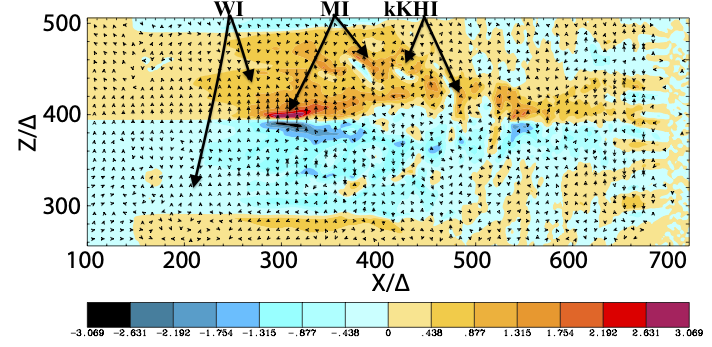}
    \end{minipage}
    \hfill
    \begin{minipage}{0.48\linewidth}
        \includegraphics[width=\linewidth]{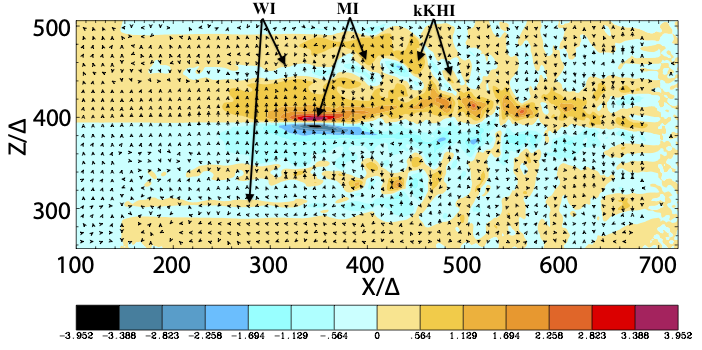}
    \end{minipage}

    \vspace{-0.06cm}

    \begin{minipage}{0.48\linewidth}
        \makebox[\linewidth][r]{\textbf{(c)}\hspace{1.0cm}}\\[-0.06cm]
        \includegraphics[width=\linewidth]{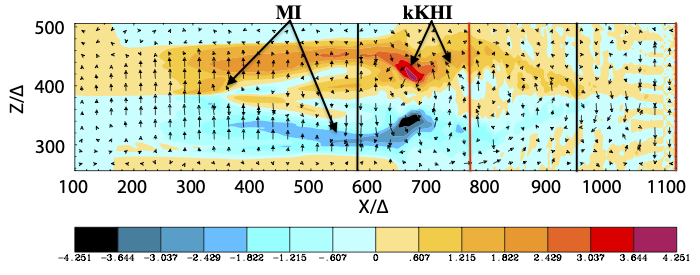}
    \end{minipage}
    \hfill
    \begin{minipage}{0.48\linewidth}
        \makebox[\linewidth][r]{\textbf{(d)}\hspace{1.0cm}}\\[-0.06cm]
        \includegraphics[width=\linewidth]{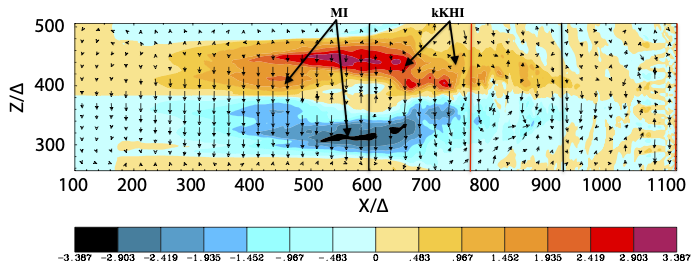}
    \end{minipage}

    \vspace{-0.06cm}

    \begin{minipage}{0.48\linewidth}
        \makebox[\linewidth][r]{\textbf{(e)}\hspace{1.0cm}}\\[-0.06cm]
        \includegraphics[width=\linewidth]{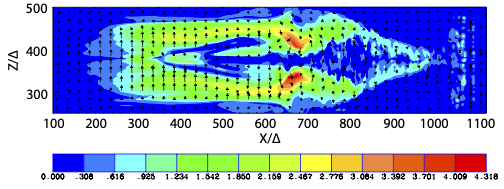}
    \end{minipage}
    \hfill
    \begin{minipage}{0.48\linewidth}
        \makebox[\linewidth][r]{\textbf{(f)}\hspace{1.0cm}}\\[-0.06cm]
        \includegraphics[width=\linewidth]{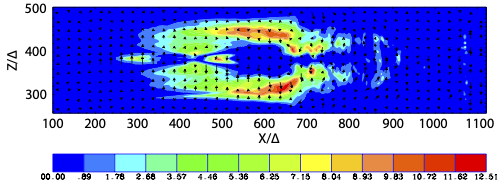}
    \end{minipage}

    \hfill

\vspace{-0.06cm}
    \caption{Evolution of the magnetic field structure for an $e^{\pm}$ jet.
    The component $B_y$ at $t = 600\,\omega_{\mathrm{pe}}^{-1}$ has extrema
    $\pm 3.069$ (a) and $\pm 3.952$ (b).
    At a later time, $t = 1000\,\omega_{\mathrm{pe}}^{-1}$, $B_y$ has extrema
    $\pm 4.251$ (c) and $\pm 3.387$ (d).
    The total magnetic field magnitude $B_{\mathrm{tot}}$ has maximum values of
    $4.318$ (e) and $12.51$ (f). Black arrows indicate the in-plane magnetic field components ( $B_x$, $B_z$).}
    \label{fig:magnetic_evolution_pa}
\end{figure*}
\begin{figure*}[t!]
    \centering

    \makebox[\linewidth]{%
        \makebox[0.46\linewidth]{\phantom{\textbf{(a)}\hspace{1.0cm}}\hfill$\gamma_{0} = 10$\hfill\textbf{(a)}\hspace{1.0cm}}%
        \makebox[0.08\linewidth][c]{\textbf{ $e^{-}$ -- $i^{+}$ jet}}%
        \makebox[0.42\linewidth]{\phantom{\textbf{(b)}\hspace{1.0cm}}\hfill$\gamma_{0} = 15$\hfill\textbf{(b)}\hspace{1.0cm}}%
    }\\[-0.06cm]

    \begin{minipage}{0.48\linewidth}
        \includegraphics[width=\linewidth]{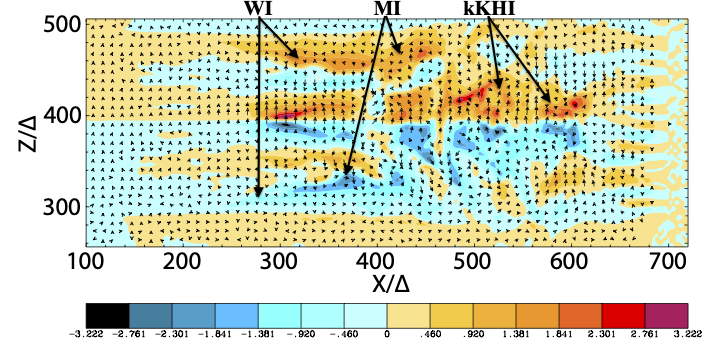}
    \end{minipage}
    \hfill
    \begin{minipage}{0.48\linewidth}
        \includegraphics[width=\linewidth]{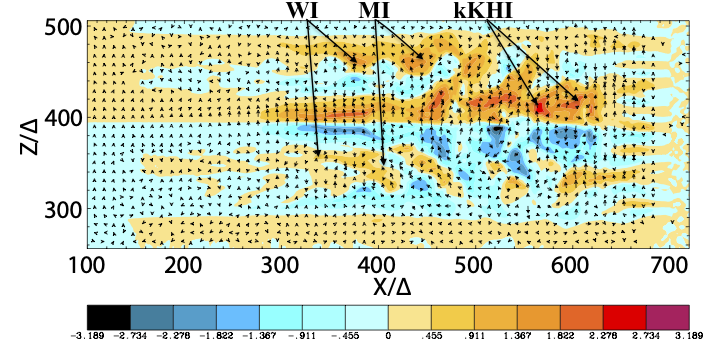}
    \end{minipage}

    \vspace{-0.06cm}

    \begin{minipage}{0.48\linewidth}
        \makebox[\linewidth][r]{\textbf{(c)}\hspace{1.0cm}}\\[-0.06cm]
        \includegraphics[width=\linewidth]{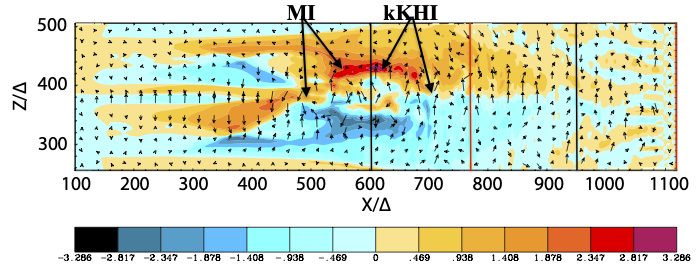}
    \end{minipage}
    \hfill
    \begin{minipage}{0.48\linewidth}
        \makebox[\linewidth][r]{\textbf{(d)}\hspace{1.0cm}}\\[-0.06cm]
        \includegraphics[width=\linewidth]{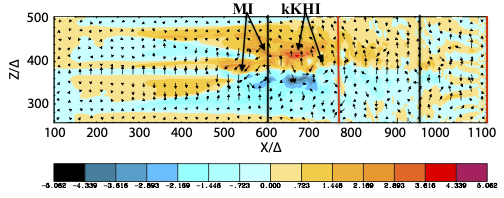}
    \end{minipage}

    \vspace{-0.06cm}

    \begin{minipage}{0.48\linewidth}
        \makebox[\linewidth][r]{\textbf{(e)}\hspace{1.0cm}}\\[-0.06cm]
        \includegraphics[width=\linewidth]{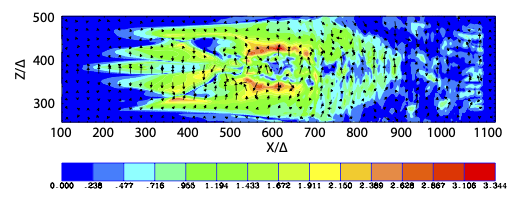}
    \end{minipage}
    \hfill
    \begin{minipage}{0.48\linewidth}
        \makebox[\linewidth][r]{\textbf{(f)}\hspace{1.0cm}}\\[-0.06cm]
        \includegraphics[width=\linewidth]{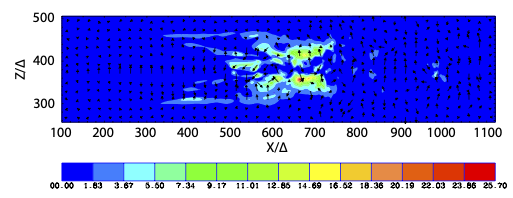}
    \end{minipage}

    \vspace{-0.06cm}
 \caption{Evolution of the magnetic field structure for an $e^{-}$ -- $i^{+}$ jet.
    The component $B_y$ at $t = 600\,\omega_{\mathrm{pe}}^{-1}$ has extrema
    $\pm 3.222$ (a) and $\pm 3.189$ (b).
    At a later time, $t = 1000\,\omega_{\mathrm{pe}}^{-1}$, $B_y$ has extrema
    $\pm 3.286$ (c) and $\pm 5.062$ (d).
    The total magnetic field magnitude $B_{\mathrm{tot}}$ has maximum values of
    $3.344$ (e) and $25.70$ (f).
    Black arrows indicate the in-plane magnetic field components ( $B_x$, $B_z$).}
    \label{fig:magnetic_evolution_ra}
\end{figure*}

Figure~\ref{fig:lorentz_electrons_combined} shows the averaged Lorentz factors of jet electrons at $t=1000\,\omega_{\mathrm{pe}}^{-1}$ for the $e^{\pm}$ jet (Figs.~\ref{fig:lorentz_electrons_combined}a and \ref{fig:lorentz_electrons_combined}c) and the $e^{-}$ -- $i^{+}$ jet (Figs.~\ref{fig:lorentz_electrons_combined}b and \ref{fig:lorentz_electrons_combined}d). In both compositions, the transverse spread of jet electrons is larger at $\gamma_0=10$ (Figs.~\ref{fig:lorentz_electrons_combined}a and \ref{fig:lorentz_electrons_combined}b) than at $\gamma_0=15$ (Figs.~\ref{fig:lorentz_electrons_combined}c and \ref{fig:lorentz_electrons_combined}d), whereas the average maximum electron Lorentz factor is higher in the $\gamma_0=15$ runs, as expected from stronger relativistic beaming. These trends are qualitatively consistent with core--sheath PIC studies showing reduced radially outward diffusion of jet electrons across the jet--ambient shear layer at larger Lorentz factors \citep{Nishikawa2014}. Although the applied toroidal field in this plane is initially zero, kinetic instabilities---specifically, the WI, MI, and kKHI---generate non-zero $x$ and $z$ components of the magnetic field near the jet core \citep{Alves2012,Alves2015,Meli2023}. Near the jet--ambient interface, the excitation of kKHI and MI is evident in both the magnetic-field evolution (as in Fig.~\ref{fig:magnetic_evolution_pa}) and in the electron Lorentz-factor maps (as in Figs.~\ref{fig:lorentz_electrons_combined} and ~\ref{fig:lorentz_positrons_ions_combined}). The two-dimensional Lorentz factor maps reveal mushroom-like (bowed layers) structures at the shear layer. We therefore interpret the transverse spreading of jet electrons as being driven primarily by the combined action of MI and kKHI. It should be noted that the structure of bowed layers of diffused jet electrons appears to be caused by the kKHI along the jet propagation, and the spacing of these layers shows the wavelength of the kKHI.

Figure~\ref{fig:lorentz_positrons_ions_combined} presents the averaged Lorentz factors of jet positrons and ions at $t = 1000\,\omega_{\mathrm{pe}}^{-1}$, showing the global jet structure for the $e^{\pm}$ jet (Figs.~\ref{fig:lorentz_positrons_ions_combined}a and ~\ref{fig:lorentz_positrons_ions_combined}c) and the $e^{-}$ -- $i^{+}$ jet (Figs.~\ref{fig:lorentz_positrons_ions_combined}b and ~\ref{fig:lorentz_positrons_ions_combined}d).
For the $e^{\pm}$ jet, the bowed layers of jet positrons are not as visible as those of jet electrons (Figs.~\ref{fig:lorentz_electrons_combined}a and ~\ref{fig:lorentz_electrons_combined}b); they appear more smeared.
For the $e^{-}$ -- $i^{+}$ jet, bowed layers are even less evident for jet ions (Figs.~\ref{fig:lorentz_positrons_ions_combined}b and ~\ref{fig:lorentz_positrons_ions_combined}d), and no clear bowed layers are seen in the ion Lorentz factor distribution. Since ions ($m_i/m_e = 4$) are heavier than positrons, ions are not effectively accelerated by the kKHI and MI.

Figures~\ref{fig:lorentz_electrons_combined} and~\ref{fig:lorentz_positrons_ions_combined} show reduced outward diffusion of jet particles at higher $\gamma_0$ for both plasma compositions. In the lower-$\gamma_0$ runs, the broader spread of jet particles arises mainly from weaker relativistic beaming combined with stronger growth of shear-driven instabilities, particularly MI, at the jet--ambient interface. This interpretation is supported by the weaker total magnetic-field strength in Figs.~\ref{fig:magnetic_evolution_pa}e and~\ref{fig:magnetic_evolution_ra}e than in Figs.~\ref{fig:magnetic_evolution_pa}f and~\ref{fig:magnetic_evolution_ra}f. The reduced outward diffusion at higher $\gamma_0$ is qualitatively consistent with core--sheath PIC simulations \citep{Nishikawa2014}.
This trend is also qualitatively consistent with the unmagnetized MI theory of \citet{Alves2015}, for which the maximum growth rate of a symmetric cold shear, $\Gamma_{\max}/\omega_{pe}=\beta_0/\sqrt{\gamma_0}$, decreases with increasing $\gamma_0$ for $\gamma_0>\sqrt{3}$. Both cases considered here, $\gamma_0=10$ and $15$, lie in this regime. Magnetized-MI analyses further indicate that stronger magnetic fields suppress the instability growth \citep{Guo2026}.

 \begin{figure*}[t!]
    \centering
    
    \begin{minipage}{0.48\textwidth}
        \makebox[\linewidth]{\phantom{\textbf{(a)}\hspace{1.0cm}}\hfill\textbf{${e^{\pm}}$ jet}\hfill\textbf{(a)}\hspace{1.0cm}}\\[-0.06cm]
        \includegraphics[width=\textwidth]{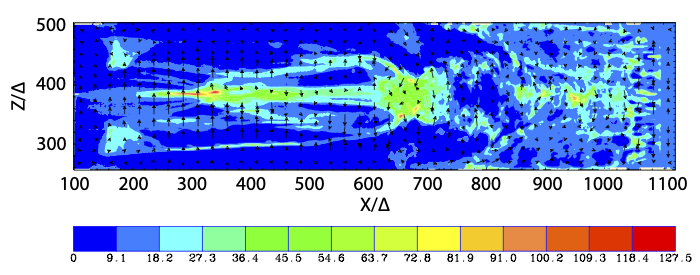}
    \end{minipage}
    \hfill
    \begin{minipage}{0.48\textwidth}
        \makebox[\linewidth]{\phantom{\textbf{(b)}\hspace{1.0cm}}\hfill\textbf{$e^{-}$ -- $i^{+}$ jet}\hfill\textbf{(b)}\hspace{1.0cm}}\\[-0.06cm]
        \includegraphics[width=\textwidth]{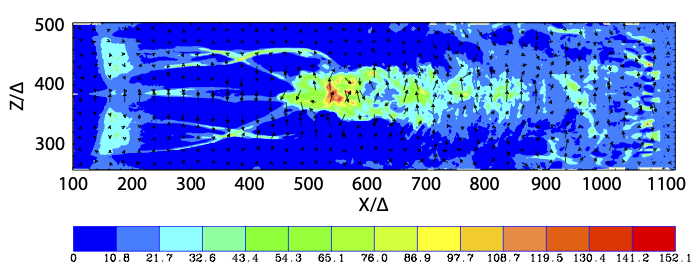}
    \end{minipage}
    \vspace{-0.06cm}
    \begin{minipage}{0.48\textwidth}
        \centering
        \noindent\makebox[\linewidth][r]{\textbf{(c)}\hspace{1.0cm}}\\
        \vspace{-0.06cm}
        \includegraphics[width=\textwidth]{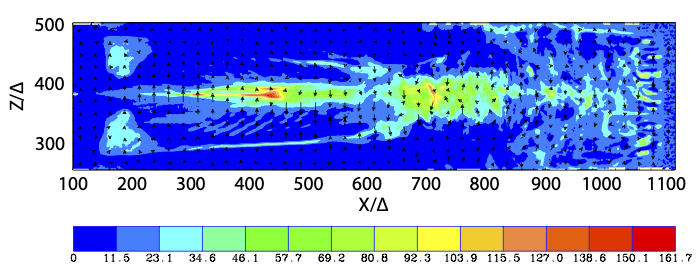}
    \end{minipage}
    \hfill
    \begin{minipage}{0.48\textwidth}
        \centering
        \noindent\makebox[\linewidth][r]{\textbf{(d)}\hspace{1.0cm}}\\
        \vspace{-0.06cm}
        \includegraphics[width=\textwidth]{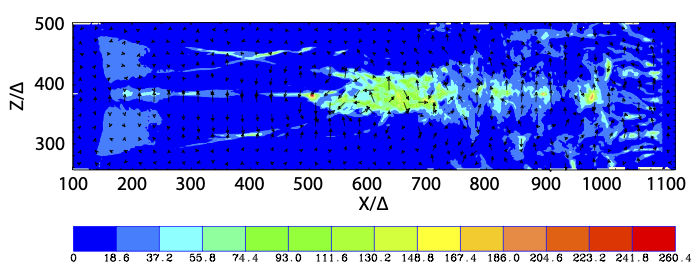}
    \end{minipage}
    \vspace{-0.06cm}
\caption{Jet electron density in the $x$--$z$ plane at
$t = 1000\,\omega_{\mathrm{pe}}^{-1}$.
Black arrows indicate the in-plane magnetic field components ($B_x$, $B_z$).
Panels (a) and (b) correspond to $\gamma_{0} = 10$, and panels (c) and (d) to
$\gamma_{0} = 15$.
Peak densities are (a) $127.5$, (b) $152.1$, (c) $161.7$, and (d) $260.4$.}
    \label{fig:density_evolution_all}
\end{figure*}
\begin{figure*}[t!]
    \centering
    \begin{minipage}{0.48\textwidth}
        \makebox[\linewidth]{\phantom{\textbf{(a)}\hspace{1.0cm}}\hfill\textbf{${e^{\pm} }$ jet}\hfill\textbf{(a)}\hspace{1.0cm}}\\[-0.06cm]
        \includegraphics[width=\textwidth]{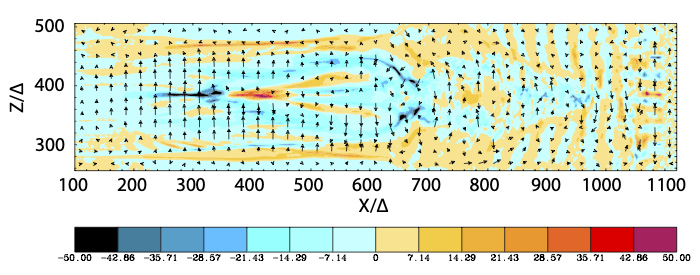}
    \end{minipage}
    \hfill
    \begin{minipage}{0.48\textwidth}
        \makebox[\linewidth]{\phantom{\textbf{(b)}\hspace{1.0cm}}\hfill\textbf{$e^{-} - i^{+}$ jet}\hfill\textbf{(b)}\hspace{1.0cm}}\\[-0.06cm]
        \includegraphics[width=\textwidth]{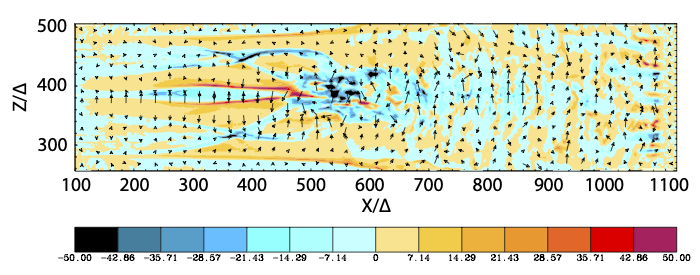}
    \end{minipage}
    \vspace{-0.06cm}
    \begin{minipage}{0.48\textwidth}
        \centering
        \noindent\makebox[\linewidth][r]{\textbf{(c)}\hspace{1.0cm}}\\
        \vspace{-0.1cm}
        \includegraphics[width=\textwidth]{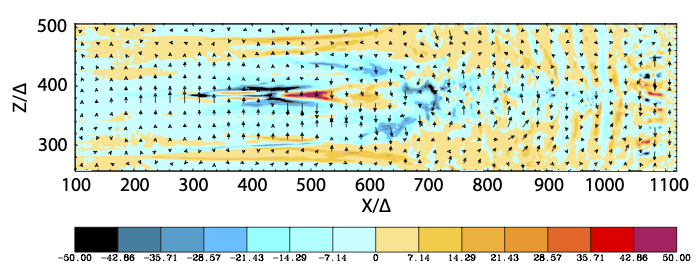}
    \end{minipage}
    \hfill
    \begin{minipage}{0.48\textwidth}
        \centering
        \noindent\makebox[\linewidth][r]{\textbf{(d)}\hspace{1.0cm}}\\
        \vspace{-0.1cm}
        \includegraphics[width=\textwidth]{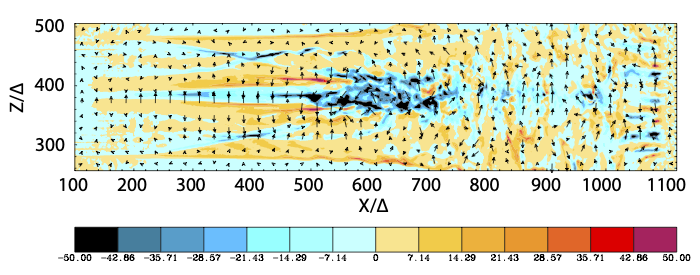}
    \end{minipage}
    \vspace{-0.1cm}
    \caption{Electric current density ($J_{x}$) in the $x$--$z$ plane at
    $t = 1000\,\omega_{\mathrm{pe}}^{-1}$, with black arrows indicating the
    in-plane magnetic field components ( $B_x$, $B_z$). For $\gamma_{0} = 10$ (a, b) and $\gamma_{0} = 15$ (c, d). In all panels, $J_{x}$ ranges between $\pm 50$.}
    \label{fig:Jx_evolution_all}
\end{figure*}


Figure~\ref{fig:magnetic_evolution_pa} presents the evolution of the magnetic field structure for an $e^{\pm}$ jet with initial Lorentz factors $\gamma_{0} = 10$ (left column) and $\gamma_{0} = 15$ (right column). The upper panels of Fig.~\ref{fig:magnetic_evolution_pa} show the amplified $B_y$ component at
$t = 600\,\omega_{\mathrm{pe}}^{-1}$ in the $x$--$z$ plane at $y/\Delta = 381$,
with the in-plane magnetic field $(B_x, B_z)$ indicated by black arrows.
In the linear regime at $t = 600\,\omega_{\mathrm{pe}}^{-1}$, the magnetic field component $B_y$ starts to grow simultaneously in the same region of the jet at the jet--ambient medium boundary and across the jet, as seen in Figs.~\ref{fig:magnetic_evolution_pa}a and~\ref{fig:magnetic_evolution_pa}b. The excitation of the MI and the kKHI merge with the WI, which results in slanted stripe structures of the magnetic field in the $e^{\pm}$ jet. Because the growth rates of MI and kKHI are similar, the excited modes propagate toward the jet center. For the higher Lorentz factor, $\gamma_{0} = 15$, the instabilities are more elongated in comparison with the lower Lorentz factor, $\gamma_{0} = 10$. In the nonlinear stage at $t = 1000\,\omega_{\mathrm{pe}}^{-1}$, the kKHI and WI become more prominent.
The labels in Figs.~\ref{fig:magnetic_evolution_pa} and~\ref{fig:magnetic_evolution_ra} are phenomenological: they mark regions where the WI, MI, and kKHI appear dominant based on characteristic morphology and spatial orientation. Because these modes coexist and couple in the nonlinear regime, the labels are not unique single-mode identifications, and the dominant locations evolve in the jet. The four supplementary \href{https://zenodo.org/records/22710808}{movies} illustrate the temporal and spatial evolution of these kinetic instabilities.

\begin{figure*}[t!]
\centering
\begin{minipage}{0.48\textwidth}
        \makebox[\linewidth]{\phantom{\textbf{(a)}\hspace{1.0cm}}\hfill\textbf{${e^{\pm} }$ jet}\hfill\textbf{(a)}\hspace{1.0cm}}\\[-0.06cm]
        \includegraphics[width=\textwidth]{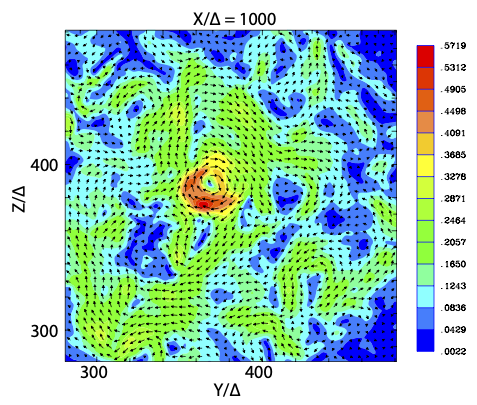}
    \end{minipage}
    \hfill
    \begin{minipage}{0.48\textwidth}
        \makebox[\linewidth]{\phantom{\textbf{(b)}\hspace{1.0cm}}\hfill\textbf{ $e^{-}$ -- $i^{+}$ jet}\hfill\textbf{(b)}\hspace{1.0cm}}\\[-0.06cm]
        \includegraphics[width=\textwidth]{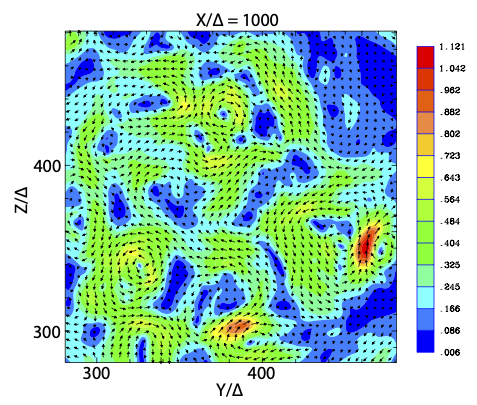}
    \end{minipage}

    \vspace{-0.06cm}

\begin{minipage}{0.48\textwidth}
  \centering
  \noindent\makebox[\linewidth][r]{\textbf{(c)}\hspace{1.0cm}}\\
  \vspace{-0.1cm}
  \includegraphics[width=\textwidth]{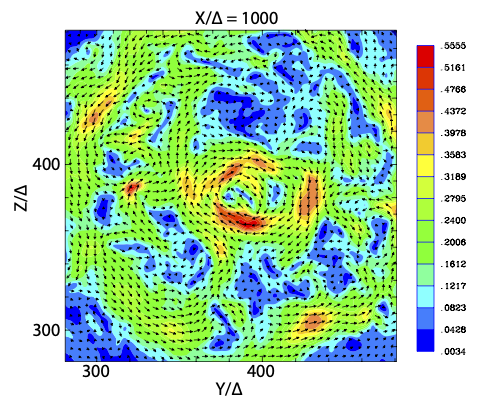}
\end{minipage}
\hfill
\begin{minipage}{0.48\textwidth}
  \centering
  \noindent\makebox[\linewidth][r]{\textbf{(d)}\hspace{1.0cm}}\\
  \vspace{-0.1cm}
  \includegraphics[width=\textwidth]{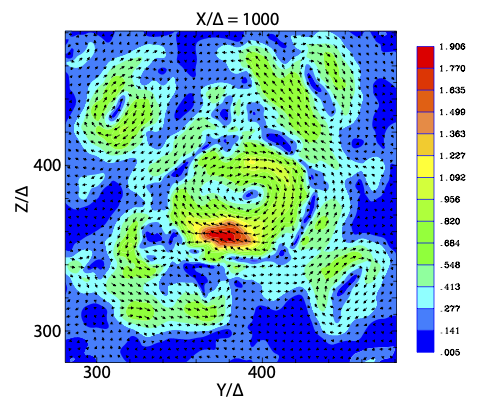}
\end{minipage}

\vspace{-0.1cm}

\caption{Total magnetic field strength in the $y$--$z$ plane at
$x/\Delta = 1000$ and $t = 1000\,\omega_{\mathrm{pe}}^{-1}$, with black arrows
indicating the in-plane magnetic field components ($B_y, B_z$) for initial
Lorentz factors of $\gamma_{0} = 10$ (a, b) and $\gamma_{0} = 15$
(c, d).}
\label{fig:Btotyzisland_combined_2by2}
\end{figure*}

\begin{figure*}[t] 
\begin{minipage}{0.48\textwidth} 
\makebox[\linewidth]{ \phantom{\textbf{(a)}\hspace{1.0cm}}
\hfill\textbf{$e^{\pm}$ jet} \hfill\textbf{(a)}\hspace{1.0cm}} \\[-0.06cm] \includegraphics[width=\textwidth] {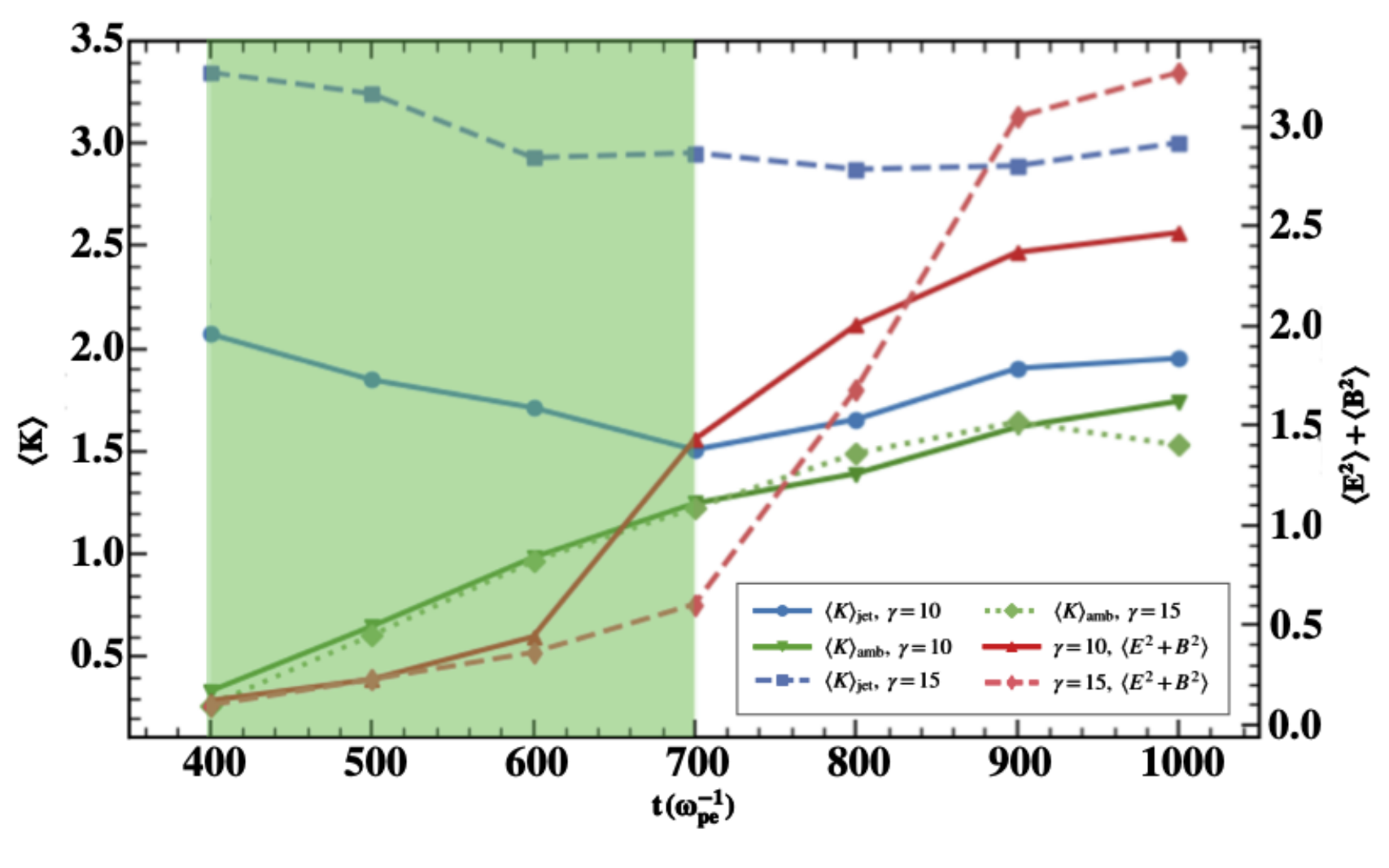} \end{minipage} \hfill \begin{minipage}{0.48\textwidth} \makebox[\linewidth]{ \phantom{\textbf{(b)}\hspace{1.0cm}}\hfill\textbf{$e^{-}$--$i^{+}$ jet}\hfill\textbf{(b)}\hspace{1.0cm}} \\[-0.06cm] \includegraphics[width=\textwidth] {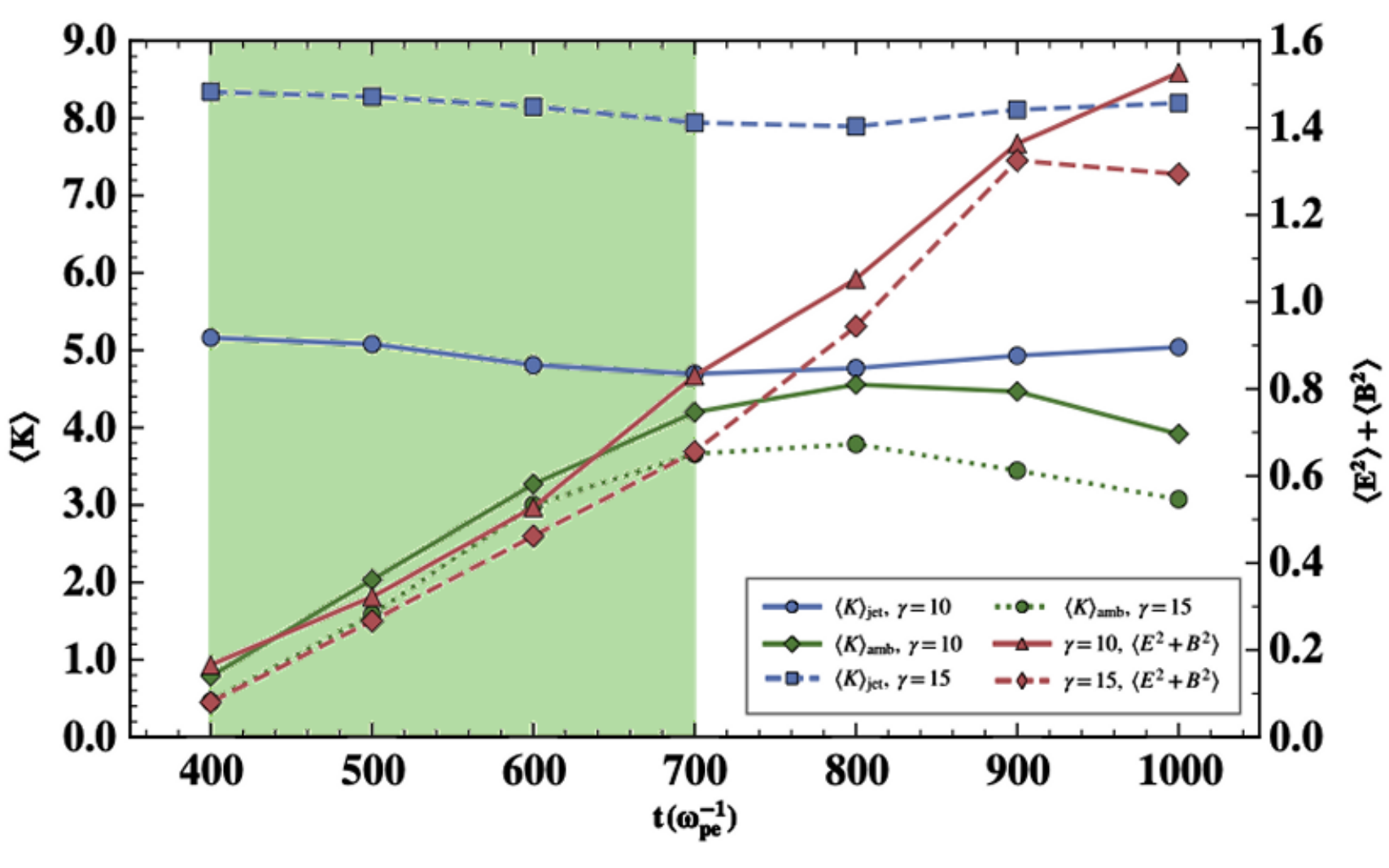} \end{minipage} 
\caption{Time evolution of the average jet particles kinetic energy($\langle K \rangle $jet), average ambient particles kinetic energy($\langle K \rangle $amb) and electromagnetic energy density($\langle E^{2}\rangle$ + $\langle  B^{2}\rangle$), measured within the volume ($x/\Delta = 300$--$650$, $y,z/\Delta = 256$--$506$) for relativistic (a) $e^{\pm}$ and (b) $e^{-}$--$i^{+}$ jets with $\gamma_{0}=10$ (solid lines) and $\gamma_{0}=15$ (dotted lines).} 
\label{fig:dEkin_em_comparison} 
\end{figure*}

As illustrated in Figs.~\ref{fig:magnetic_evolution_pa}e and  ~\ref{fig:magnetic_evolution_pa}f,
the total magnetic field distributions reveal a significant dependence on the initial Lorentz factor ($\gamma_{0}$) as expected.
For the $e^{\pm}$ jet with $\gamma_{0} = 10$, the maximum total magnetic field
is $4.318$ (as shown in Fig.~\ref{fig:magnetic_evolution_pa}e), representing an
amplification factor of approximately $17.2$ relative to the peak value of initial field
strength.
In contrast, the higher-$\gamma_{0}$ jet ($\gamma_{0} = 15$) exhibits
significantly stronger magnetic field amplification, reaching a peak magnitude
of $12.51$ (as shown in Fig.~\ref{fig:magnetic_evolution_pa}f), corresponding
to a 50-fold amplification.
While the transverse $B_y$ component remains slightly larger in the
lower-$\gamma_{0}$ case (see Figs.~\ref{fig:magnetic_evolution_pa}c and~\ref{fig:magnetic_evolution_pa}d
), the 50-fold increase in the total field strength for
$\gamma_{0} = 15$ confirms a vigorous development of the longitudinal ($B_x$)
and transverse ($B_z$) field components. The stronger total magnetic field in the higher-$\gamma_{0}$ case suppresses the growth of the mushroom instability (MI), thereby reducing the transverse (radial) spreading of jet particles (Figs.~\ref{fig:lorentz_electrons_combined}c, d and \ref{fig:lorentz_positrons_ions_combined}c, d) relative to the lower-$\gamma_0$ case (Figs.~\ref{fig:lorentz_electrons_combined}a, b and \ref{fig:lorentz_positrons_ions_combined}a, b). In contrast, the lower-$\gamma_0$ case exhibits stronger MI development and consequently greater transverse spreading of the jet particles. The kKHI develops primarily along the longitudinal jet–ambient shear layer, distinguishing its characteristic spatial development from the transverse MI structures. The supplementary movies further illustrate the temporal and spatial development of the MI and kKHI along the jet. Together, these results suggest that while higher $\gamma_{0}$ governs the
growth and ultimate saturation of the total field, the lower-$\gamma_{0}$ case
favors more rapid development of kinetic instabilities.

Similarly, for the $e^{-}$ -- $i^{+}$ jet, the magnetic field structure is shown in Fig.~\ref{fig:magnetic_evolution_ra}. For the higher Lorentz factor, $\gamma_{0} = 15$, the peak value
of the total magnetic field is $25.70$ (as shown in Fig.~\ref{fig:magnetic_evolution_ra}f), approximately 103 times the initial peak value of magnetic field. This sharp amplification indicates that higher Lorentz factors lead to higher peak magnetic-field amplitudes and a more complex three-dimensional field topology at $t = 1000\,\omega_{pe}^{-1}$, in which the initial flow energy directly sets the saturation level of the magnetic field.
This aspect is shown in Fig. \ref{fig:dEkin_em_comparison} which will be explained later. These results indicate that $\gamma_{0}$ significantly influences instability
growth and particle energization.
Moreover, at higher $\gamma_{0}$, the peak value of the total magnetic field is greater for the $e^{-}$ -- $i^{+}$ jet
compared with those of $e^{\pm}$ jets.
This suggests that plasma composition contributes significantly to the nonlinear
evolution of the instabilities.

For both jets with different compositions, the magnetic field filaments begin to grow simultaneously at
the jet--ambient medium boundary and across the jet intThe serior, as seen in
Figs.~\ref{fig:magnetic_evolution_pa}a, b and
\ref{fig:magnetic_evolution_ra}a, b.
The excitation of MI and kKHI couples with the WI,
producing slanted stripe patterns in the magnetic field of the $e^{\pm}$ jet.
Because the growth rates of MI and kKHI are similar in this regime, the excited
modes propagate toward the jet center.
In both the $e^{\pm}$ and $e^{-}$ -- $i^{+}$ jet cases, the structures observed
at the jet--ambient plasma boundary in the region $500 < x/\Delta < 800$
indicate the excitation and evolution of the kKHI and MI. Notably, within the range $500 < x/\Delta < 700$, the reversal of the
magnetic field is visible at the jet center (as shown in
Figs.~\ref{fig:magnetic_evolution_pa}c, d and \ref{fig:magnetic_evolution_ra}c, d),
accompanied by a localized deceleration of the flow.
This leads to the formation of a narrow throat-like structure, reflecting strong
jet-electron collimation (as seen in Fig.~\ref{fig:density_evolution_all}). Figures~\ref{fig:density_evolution_all}a, c and~\ref{fig:density_evolution_all}b, d show the electron density colormap for $e^{\pm}$ and $e^{-}$ -- $i^{+}$ jets at $t = 1000\,\omega_{\mathrm{pe}}^{-1}$ for $\gamma_{0}=10$ (upper panels) and $\gamma_{0}=15$ (lower panels), respectively.
The enhanced collimation of jet electrons results from the narrow jet geometry
and significantly suppressed growth of MI due to amplified total magnetic field as shown in Figs.~\ref{fig:magnetic_evolution_pa}e, f and ~\ref{fig:magnetic_evolution_ra}e, f.
At $t = 1000\,\omega_{\mathrm{pe}}^{-1}$, the collimation of the electron beams
with $\gamma_{0} = 15$ begins earlier than for the lower-$\gamma_{0}$ case,
consistent with the current density $J_x$ shown in 
Fig.~\ref{fig:Jx_evolution_all}c. This is due to the combined effect of the stronger kKHI in the longitudinal direction and weaker growth of the MI in the transverse direction.
For $\gamma_{0} = 10$, the electron density structure resembles a dumbbell,
while for $\gamma_{0} = 15$ it resembles a bottleneck. 
Furthermore, jet electrons show concentration at smaller spatial scales for
$\gamma_{0} = 15$.
For $\gamma_{0} = 10$, the $e^{\pm}$ jet in
Fig.~\ref{fig:density_evolution_all}a remains relatively smooth
along $x/\Delta$, with prominent high-density features near
$x/\Delta \approx 320$ and $700$; field vectors are broadly aligned with the
flow with modest local variation.
The $e^{-}$ -- $i^{+}$ jet at the same $\gamma_{0}$
in Fig.~\ref{fig:density_evolution_all}b is more fragmented, with
stronger small-scale density contrast and a higher peak density in comparison with the
$e^{\pm}$ jet.
For $\gamma_{0} = 15$, the collimation of jet electrons and peak density in the
$e^{\pm}$ jet (Fig.~\ref{fig:density_evolution_all}c) are greater
relative to those at $\gamma_{0} = 10$
(Fig.~\ref{fig:density_evolution_all}a).
The $e^{-}$ -- $i^{+}$ jet at $\gamma_{0} = 15$
(Fig.~\ref{fig:density_evolution_all}d) is the most disrupted:
multiple filaments and intense compressions appear between
$x/\Delta \approx 600$--$900$, reaching the largest peak density among the four
cases.
In all cases, the in-plane magnetic field components ($B_x$, $B_z$) are
compressed and twisted in regions of enhanced density, consistent with
magnetized flow evolution within the jet interior, shear layer, and turbulent
envelope.
The strongest collimation of jet electrons is observed in the
 $e^{-}$ -- $i^{+}$ jet with $\gamma_{0} = 15$, which exhibits the greatest peak
values of both Lorentz factor (see Fig.~\ref{fig:lorentz_electrons_combined}d) and total magnetic field (see Fig.~\ref{fig:magnetic_evolution_ra}f) among the four cases.

For the $e^{\pm}$ jet with $\gamma_{0} = 10$, a strong positive current density
$J_x$ develops in the jet's central spine region ($350 < x/\Delta < 450$) owing
to merging inner MI modes, while a negative $J_x$ component emerges near
$x/\Delta \approx 680$ as outer MI and kKHI modes propagate inward, creating
alternating regions of concentrated current.
The structural differences in the electron density and current profiles (see Figs.~\ref{fig:density_evolution_all} and \ref{fig:Jx_evolution_all}) indicate that $\gamma_{0}$ directly affects the jet morphology and long-term stability,
and plays a key role in shaping the large-scale structure of AGN jets.
Furthermore, the plasma composition largely determines the extent of jet-electron
collimation.

Figure~\ref{fig:Btotyzisland_combined_2by2} depicts the total magnetic field
structure in the $y$--$z$ plane at $x/\Delta = 1000$ for the $e^{\pm}$ jet
(Figs.~\ref{fig:Btotyzisland_combined_2by2}a and~\ref{fig:Btotyzisland_combined_2by2}c) and the $e^{-}$ -- $i^{+}$ jet (Figs.~\ref{fig:Btotyzisland_combined_2by2}b and~\ref{fig:Btotyzisland_combined_2by2}d) at $t = 1000\,\omega_{\mathrm{pe}}^{-1}$,
respectively.
In both cases, the $y$--$z$ cross-sections reveal a highly turbulent magnetic
field morphology during the late linear stage.
A spatial analysis of the magnetic field at this stage shows numerous magnetic
islands with randomly oriented magnetic field vectors, consistent with the
observed diffusion of jet electrons following jet collimation.
Most importantly, these results indicate that once particles are injected, they
undergo further acceleration via stochastic scattering off turbulent
fluctuations. From the particle energy distribution curves in Fig.~\ref{fig:combined_energy_dist}, we find that the jet electrons are further accelerated at the magnetic field dissipation occurred in the nonlinear stage \citep{Meli2023}, as also observed in kinetic simulations of driven magnetized turbulence \citep[e.g.,][]{Zhdankin2018, Bresci2023}. We note that electrons can be further accelerated to higher Lorentz factors via
turbulent acceleration, as demonstrated in kinetic simulations \citep[e.g.,][]{Comisso2019, Zhdankin2019}.
In those studies, turbulent magnetic fluctuations are externally driven; thus,
the energy source for the turbulence is not self-consistent.
In contrast, our simulations employ turbulent magnetic
fields---manifesting as multiple magnetic islands as shown in
Fig.~\ref{fig:Btotyzisland_combined_2by2}---that are self-consistently
generated in the relativistic jet through the relaxation of the toroidal
magnetic field.
Consequently, particles can be directly accelerated both within the turbulent
magnetic field regions and through interactions with the magnetic islands
visible in Fig.~\ref{fig:Btotyzisland_combined_2by2} as in \citep{Meli2023}.

\subsection{Electromagnetic Field and Particle Acceleration}
\label{sec:acc}

Figure~\ref{fig:Ex_phase_4x2_combined} shows the electric field component
$E_{\rm x}$ and the corresponding phase-space distributions ($x$--$\gamma v_{x}$) of jet and ambient electrons for $e^{\pm}$ and $e^{-}$ -- $i^{+}$ jets with
$\gamma_{0} = 10$ (left column) and $\gamma_{0} = 15$ (right column). 
The electric-field stripes are more slanted (oblique) in the
$e^{-}$ -- $i^{+}$ jet than in the $e^{\pm}$ jet, consistent with stronger and
more prominent instability growth.
Because the growth rates of MI and kKHI are comparable, the excited modes
propagate toward the jet center and shape the developing kinetic
instabilities.
In both cases, quasi-steady electric fields accelerate jet electrons in
multiple bunches due to the excitation of kKHI with MI. 
	Figure~\ref{fig:dEkin_em_comparison} shows the time evolution of the electromagnetic energy density, $\langle E^{2}\rangle+\langle B^{2}\rangle$, together with the average kinetic energies of the jet and ambient particles. During the shaded interval, the electromagnetic energy density and the average kinetic energy of the ambient particles increase, while the average kinetic energy of the jet particles decreases. This concurrent evolution is consistent with the transfer of jet kinetic energy to electromagnetic fields and ambient particles through the developing kinetic instabilities. During earlier time interval, $t=(400$--$700)\omega_{\mathrm{pe}}^{-1}$ (shaded region in Figure~\ref{fig:dEkin_em_comparison}), the electromagnetic energy density grows more rapidly for $\gamma_0=10$ than for $\gamma_0=15$ in both the pair and electron--ion jets, indicating stronger early-time instability-driven field growth at the lower Lorentz factor. In Fig.~\ref{fig:dEkin_em_comparison}a, the electromagnetic energy density  of the $e^{\pm}$ jet for the $\gamma_0=10$ (solid red curve) initially grows more rapidly, whereas that for the $\gamma_0=15$ (dashed red curve) exceeds it after $t\simeq850\omega_{\mathrm{pe}}^{-1}$.

Figure~\ref{fig:Ex_phasepa_combined_all} shows the electric field structure and
phase-space distributions, revealing possible acceleration mechanisms for jet
electrons, positrons, and ambient electrons for the $e^{\pm}$ jet in the nonlinear stage at $t = 1000\omega_{\mathrm{pe}}^{-1}$. 
Figures~\ref{fig:Ex_phasepa_combined_all}c and~\ref{fig:Ex_phasepa_combined_all}d show the
phase-space distributions ($x$--$\gamma v_x$ ) of jet electrons, ambient electrons,
and positrons for $\gamma_{0} = 10$ (left) and $\gamma_{0} = 15$ (right),
respectively. We sampled a subset of jet and ambient particles for diagnostics to avoid excessively large datasets; therefore, the phase-space plots are approximate. The peak of jet-electron acceleration (red dots) occurs at $x/\Delta = 990$
	for $\gamma_0 = 10$ and at $x/\Delta = 1000$ for $\gamma_0 = 15$.
	For electrons with $v_x>0$, regions of $E_x<0$ accelerate the electrons
	along $+x$, whereas regions of $E_x>0$ decelerate them.
	This alternating longitudinal electric field (see Figs.~\ref{fig:Ex_phasepa_combined_all}a, b and
	\ref{fig:Ex_phas_ra_combined_comparison}a, b)
	creates the observed spatial pattern of electron acceleration and deceleration
	behind the jet head (see Figs.~\ref{fig:Ex_phasepa_combined_all}c, d and
	\ref{fig:Ex_phas_ra_combined_comparison}c, d).

\begin{figure*}[t!]
\centering
    \makebox[\linewidth]{%
        \makebox[0.46\linewidth]{\phantom{\textbf{(a)}\hspace{1.0cm}}\hfill$\gamma_{0} = 10$\hfill\textbf{(a)}\hspace{1.0cm}}%
         \makebox[0.1\linewidth][c]{\textbf{$e^{\pm}$ jet}}%
        \makebox[0.46\linewidth]{\phantom{\textbf{(b)}\hspace{1.0cm}}\hfill$\gamma_{0} = 15$\hfill\textbf{(b)}\hspace{1.0cm}}
    }\\[0.2cm]
\begin{minipage}{0.46\textwidth}
   \includegraphics[width=\textwidth]{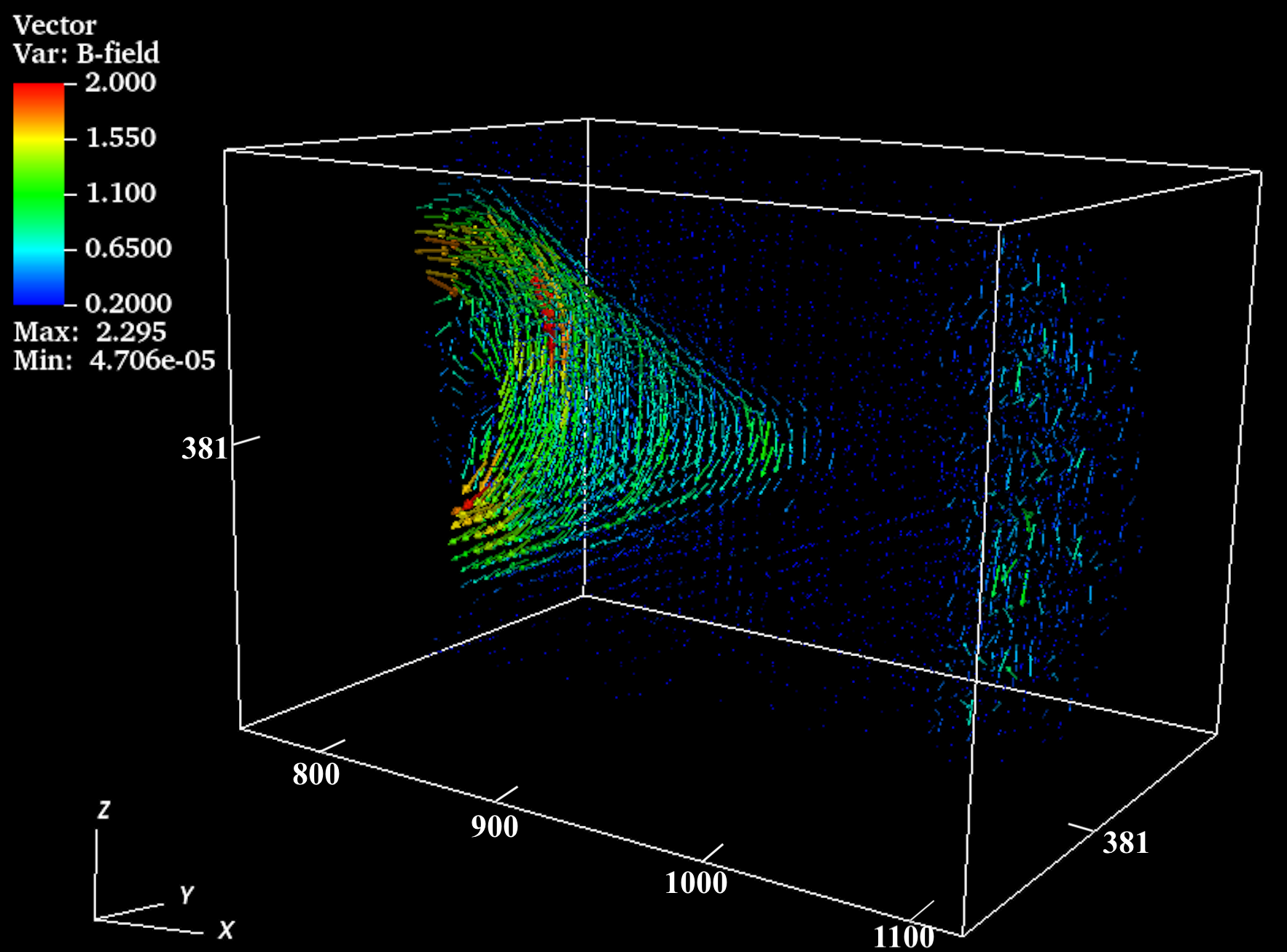}
\end{minipage}
\hfill
\begin{minipage}{0.46\textwidth}
  \includegraphics[width=\textwidth]{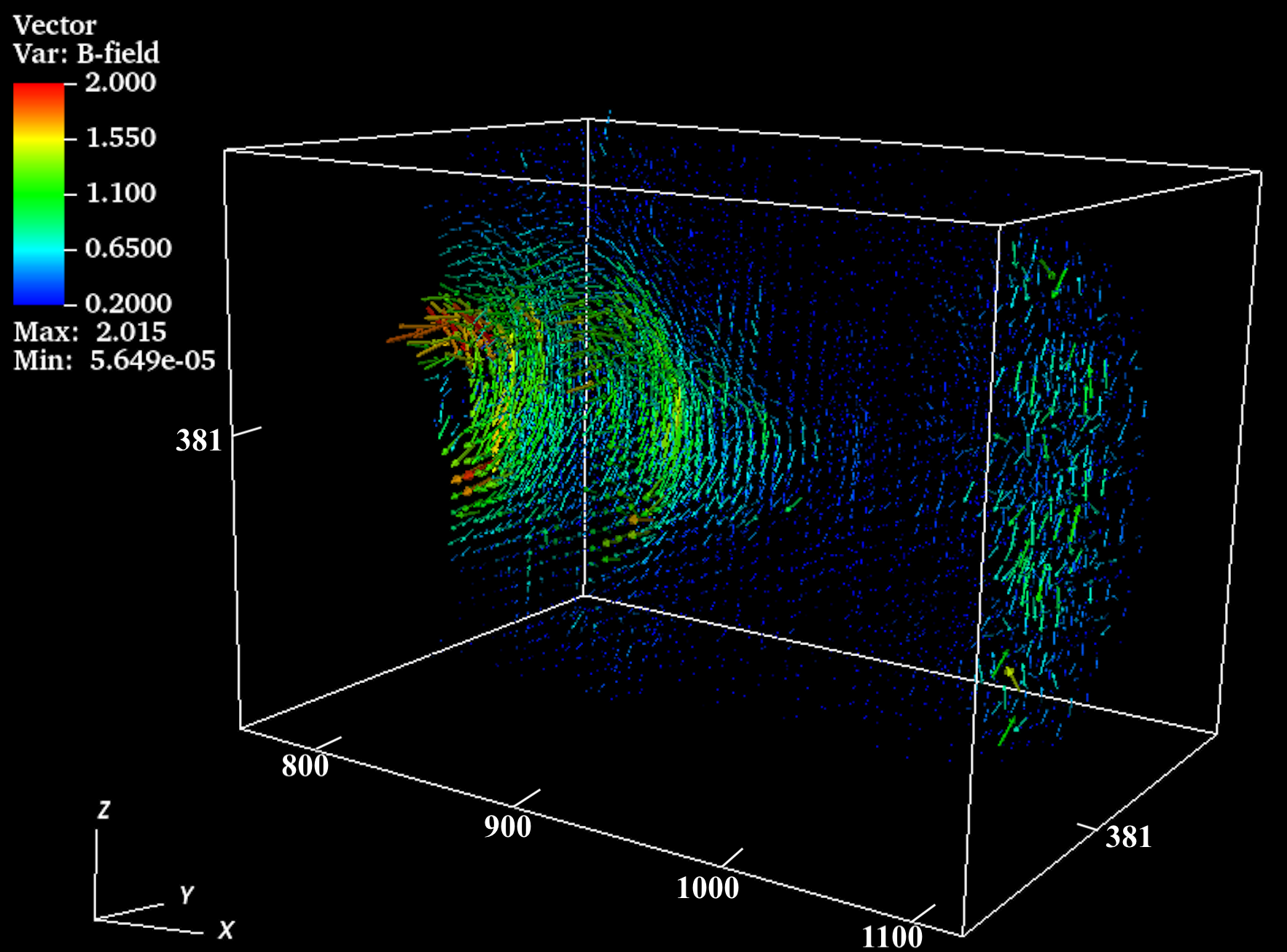}
\end{minipage}
\vspace{0.3cm}
    \makebox[\linewidth]{%
        \makebox[0.46\linewidth]{\phantom{\textbf{(c)}\hspace{1.0cm}}\hfill\textbf{(c)}\hspace{1.0cm}}%
        \makebox[0.1\linewidth][c]{\textbf{ $e^{-}$ -- $i^{+}$ jet}}%
        \makebox[0.46\linewidth]{\phantom{\textbf{(d)}\hspace{1.0cm}}\hfill\textbf{(d)}\hspace{1.0cm}}%
    }\\[-0.1cm]
\begin{minipage}{0.46\textwidth}
  \includegraphics[width=\textwidth]{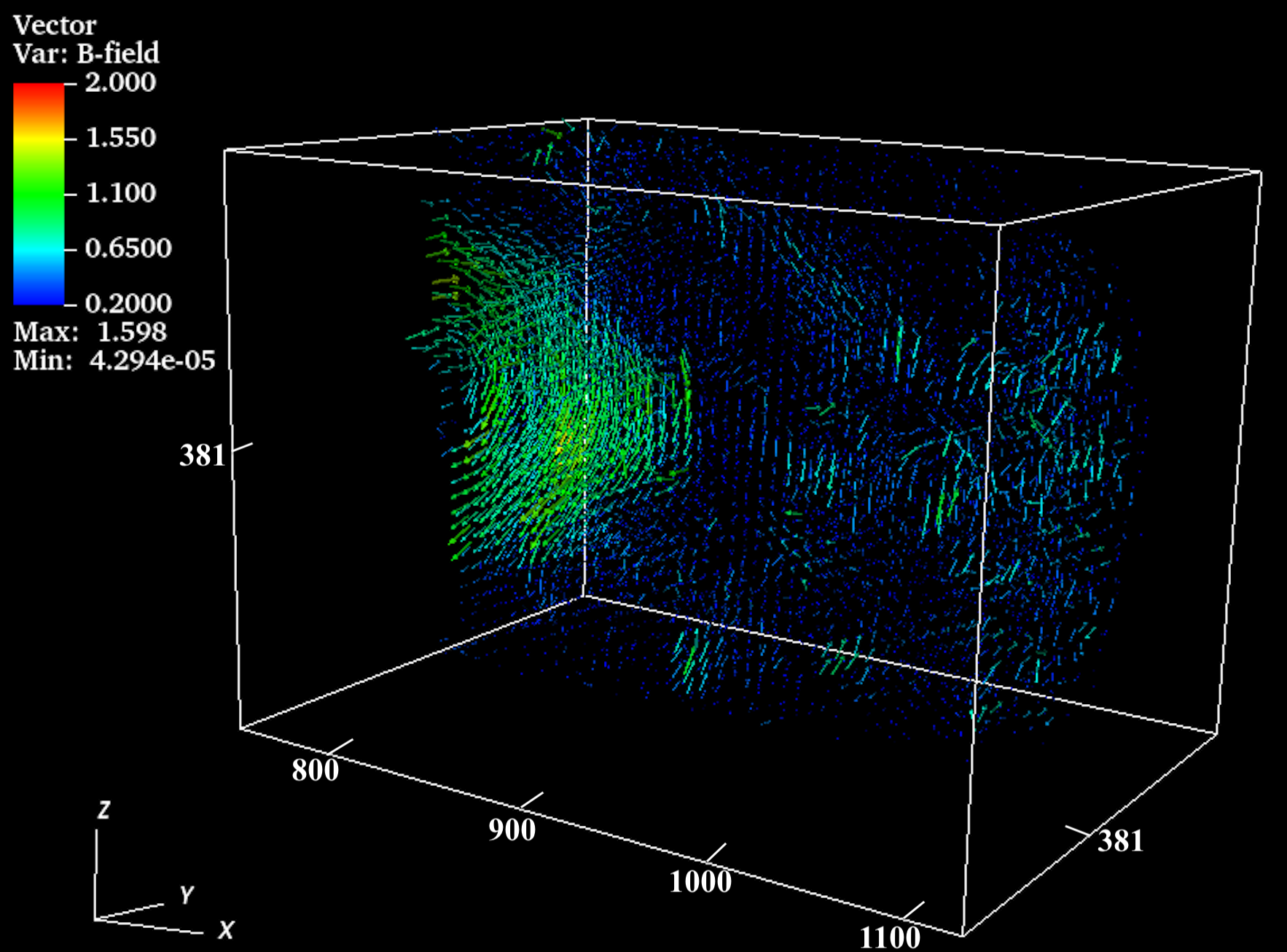 }
\end{minipage}
\hfill
\begin{minipage}{0.46\textwidth}
  \includegraphics[width=\textwidth]{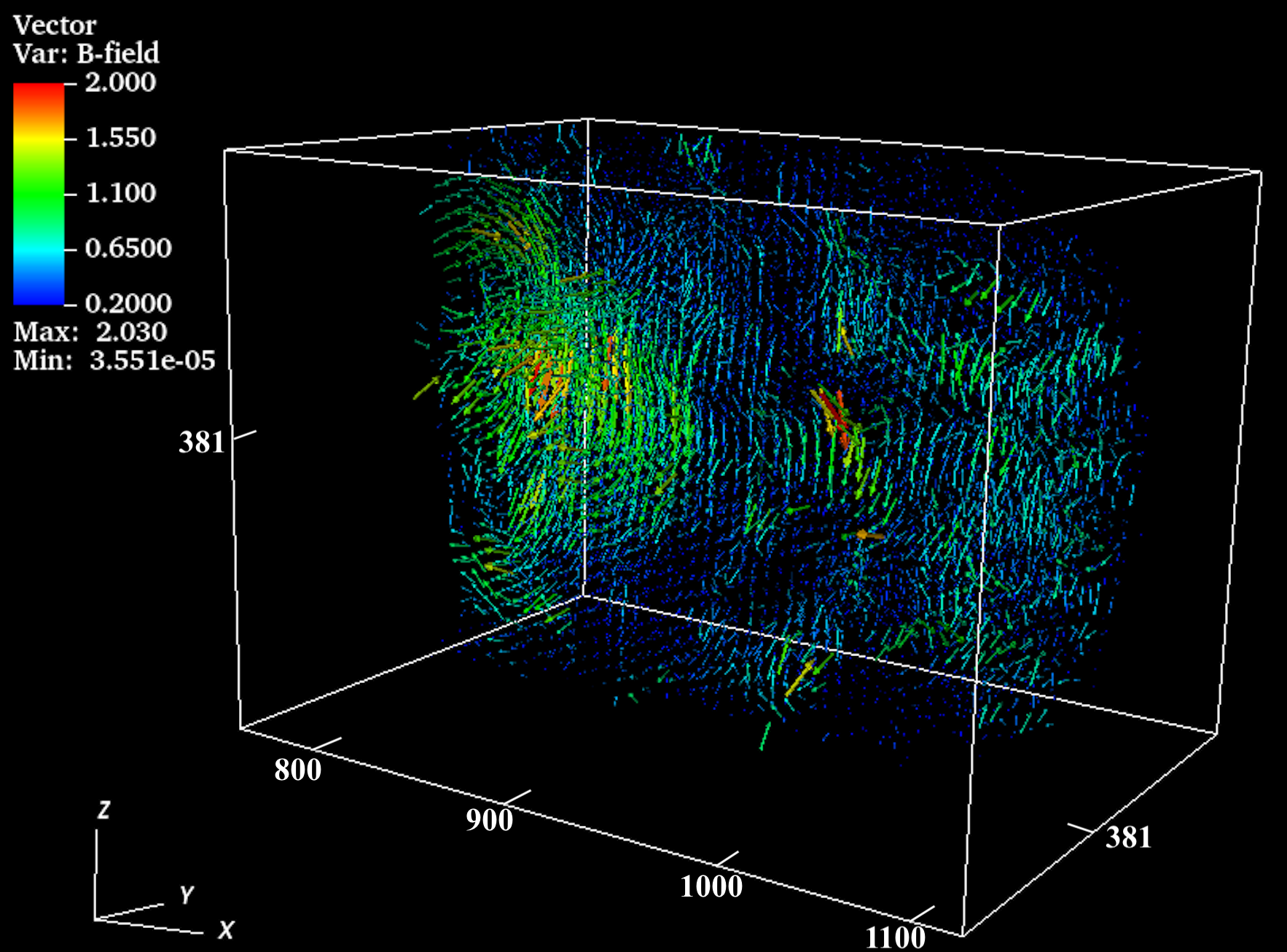}
\end{minipage}
\caption{Magnetic field vectors within a cuboid volume
($770 < x/\Delta < 1120$; $256 < y/\Delta,\, z/\Delta < 506$) at
$t = 1000\,\omega_{\mathrm{pe}}^{-1}$.
The jet center is located at $y/\Delta = z/\Delta = 381$. The plots show half-regions clipped at the jet axis in
the $x$--$z$ plane ($381 < y/\Delta < 506$).}
\label{fig:3D_field_evolution_combined}
\end{figure*}

Figure~\ref{fig:Ex_phas_ra_combined_comparison} shows the electric field structure and
phase-space distributions for the $e^{-}$ -- $i^{+}$ jet. The electric field generated in the $e^{-}$ -- $i^{+}$ jet is greater than in the $e^{\pm}$ jet, whereas ion acceleration and deceleration are weaker (see Figs.~\ref{fig:Ex_phas_ra_combined_comparison}c, d) than for positrons because of the larger ion mass.
In both $e^{\pm}$ and $e^{-}$ -- $i^{+}$ jets, the acceleration in the linear stage is due to the quasi-steady
$E_{\rm x}$ stripes generated by kinetic instabilities; these stripes become more
vertically oriented in the nonlinear stage.
We find a clear correspondence between the energy gains and losses of ambient
and jet electrons.
Jet electrons remain phase-locked to saturated instabilities and produce sharp,
localized peaks in the $x-\gamma v_x$ distribution.
The ambient electrons are decoupled from these waves and exhibit broader,
inclined acceleration peaks that trail the jet front.
Importantly, ambient-electron acceleration halts at $x/\Delta \approx 880$,
which coincides with the dissipation of electromagnetic fields near
$x/\Delta \approx 800$.
Moreover, particle acceleration in the nonlinear stage is not solely due to the
electric field generated by the growth of instabilities.
The total magnetic field at $t = 1000\,\omega_{\mathrm{pe}}^{-1}$ and
$x/\Delta = 1000$ reveals a turbulent structure with multiple magnetic islands of oppositely directed, swirling magnetic field, suggesting possible reconnection-like events, which require further investigation by analyzing critical points with the Jacobian matrix \citep[e.g.,][]{Cai2007}. At $t = 1000\,\omega_{\mathrm{pe}}^{-1}$, the ambient electron acceleration in
the $e^{\pm}$ jet (Figs.~\ref{fig:Ex_phasepa_combined_all}c, d and
\ref{fig:Ex_phas_ra_combined_comparison}c, d) spans $800 < x/\Delta < 1050$,
coinciding with the jet region where strong magnetic turbulence persists after
field dissipation (Figs.~\ref{fig:magnetic_evolution_pa}e, f and \ref{fig:magnetic_evolution_ra}e, f).
Jet electrons are strongly pinched and decelerated in $250 < x/\Delta < 550$,
then gradually reaccelerated as stratifications form via the
kKHI and MI.
Some ambient electrons also attain high energies as they are swept into the
relativistic jet plasma.

\subsection{3D Magnetic Field Evolution}

\begin{figure*}[t!]
\centering

    \makebox[\linewidth]{%
        \makebox[0.46\linewidth]{\phantom{\textbf{(a)}\hspace{1.0cm}}\hfill$\gamma_{0} = 10$\hfill\textbf{(a)}\hspace{1.0cm}}%
         \makebox[0.1\linewidth][c]{\textbf{$e^{\pm}$ jet}}%
        \makebox[0.46\linewidth]{\phantom{\textbf{(b)}\hspace{1.0cm}}\hfill$\gamma_{0} = 15$\hfill\textbf{(b)}\hspace{1.0cm}}
    }\\[0.2cm]
\begin{minipage}{0.46\textwidth}
  \includegraphics[width=\textwidth]{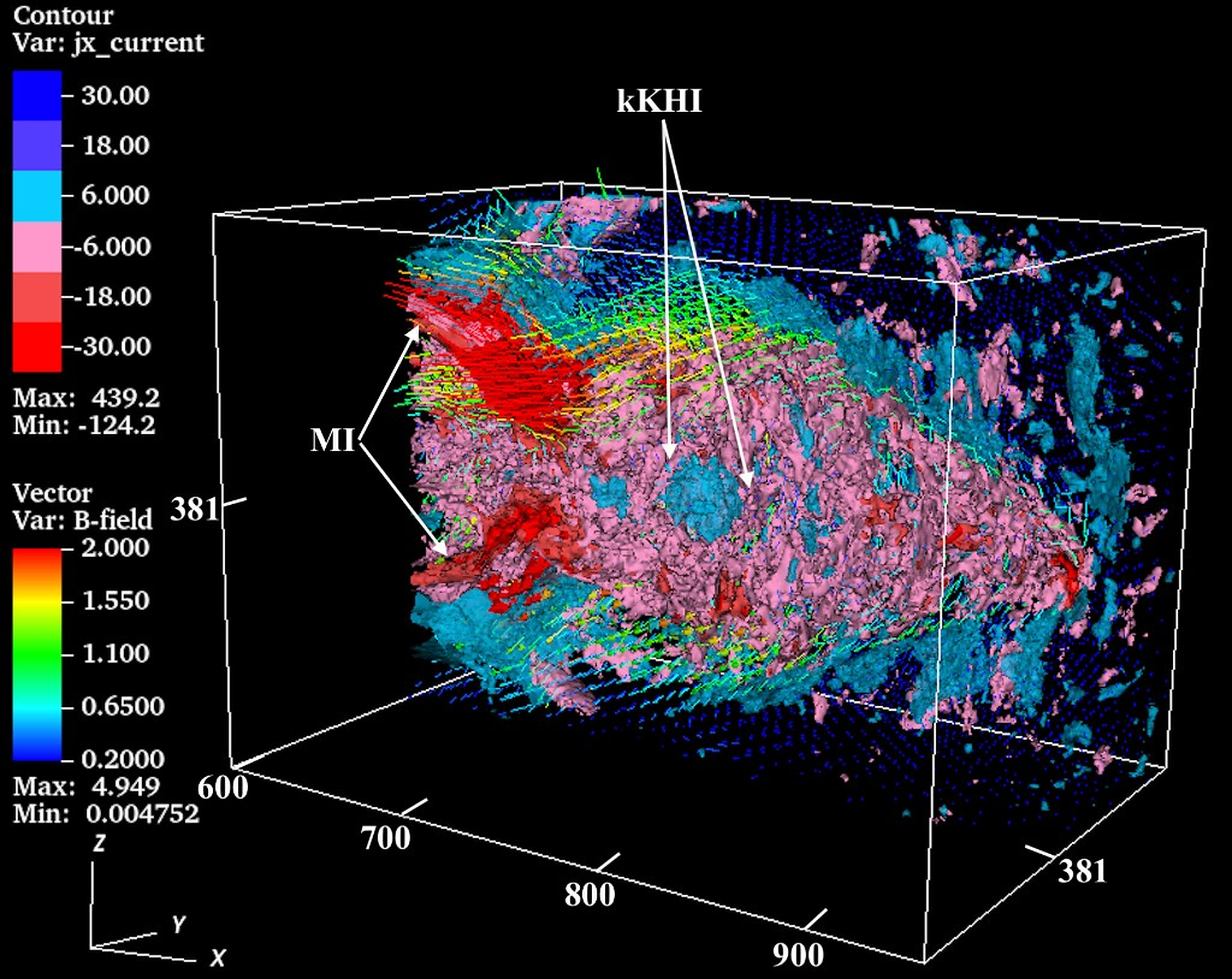}
\end{minipage}
\hfill
\begin{minipage}{0.46\textwidth}
  \includegraphics[width=\textwidth]{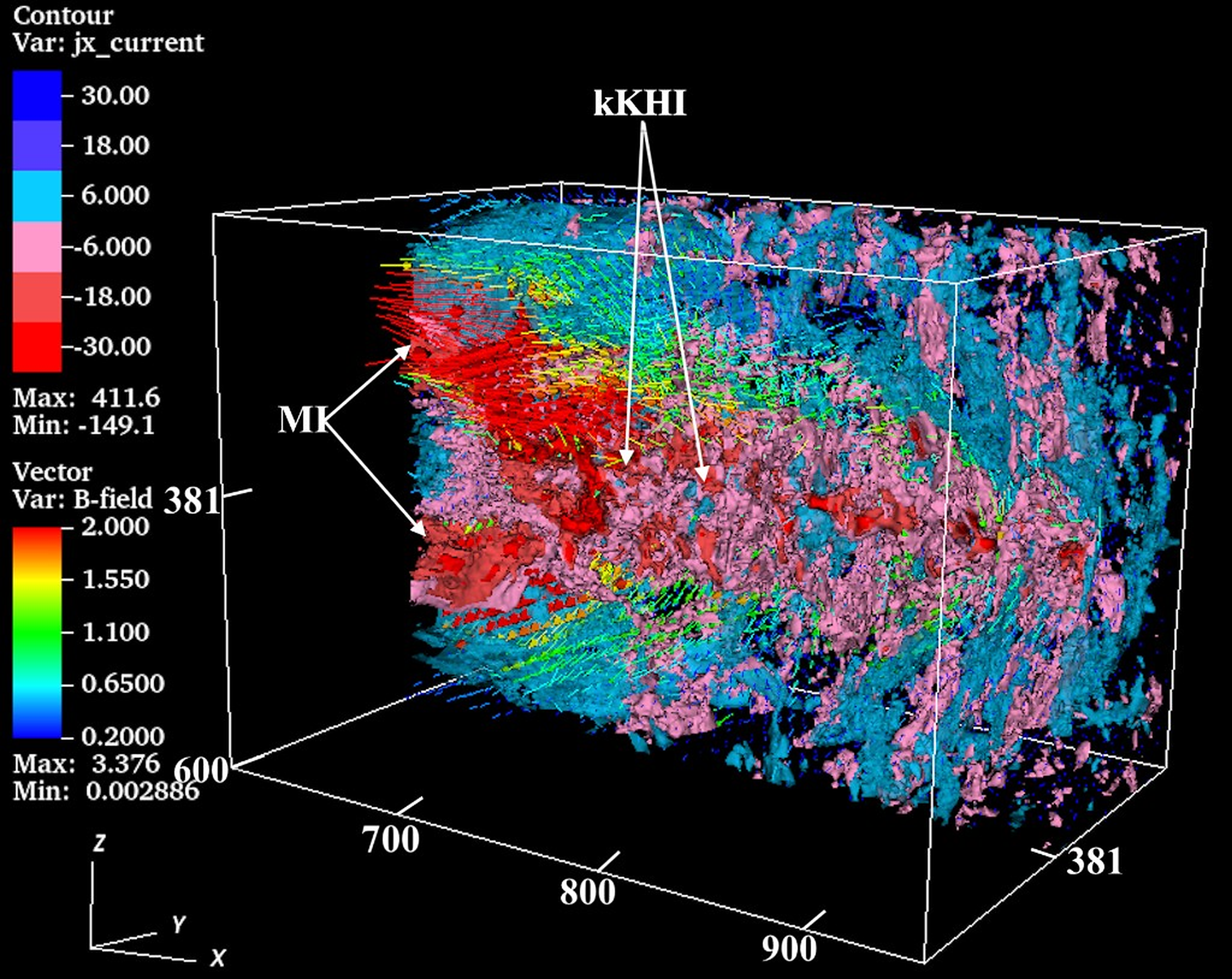}
\end{minipage}
\vspace{0.2cm}
    \makebox[\linewidth]{%
        \makebox[0.46\linewidth]{\phantom{\textbf{(c)}\hspace{1.0cm}}\hfill\textbf{(c)}\hspace{1.0cm}}%
        \makebox[0.12\linewidth][c]{\textbf{$e^{-}$ -- $i^{+}$ jet}}%
        \makebox[0.42\linewidth]{\phantom{\textbf{(d)}\hspace{1.0cm}}\hfill\textbf{(d)}\hspace{1.0cm}}%
    }\\[-0.06cm]
\begin{minipage}{0.46\textwidth}
  \includegraphics[width=\textwidth]{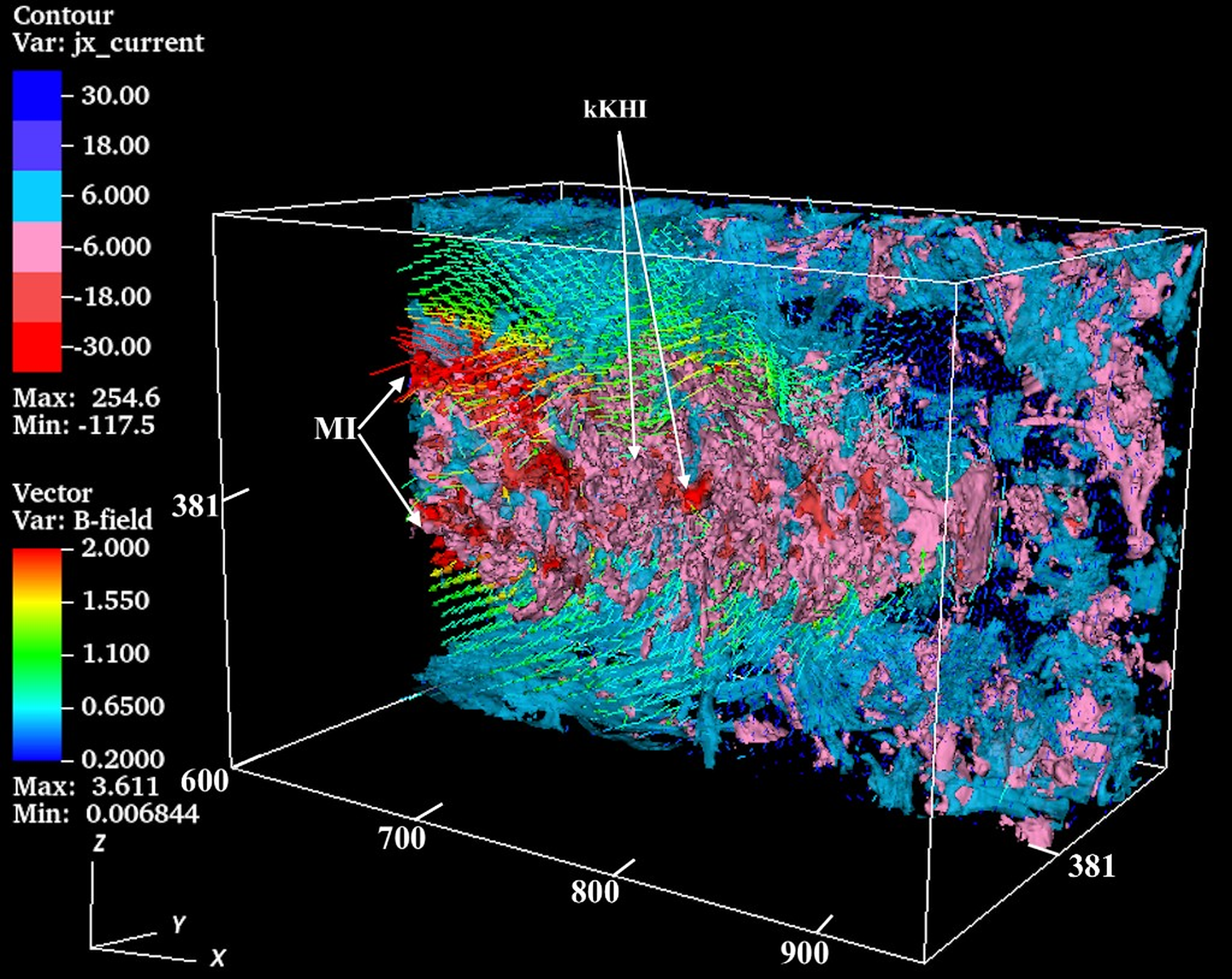}
\end{minipage}
\hfill
\begin{minipage}{0.46\textwidth}
  \includegraphics[width=\textwidth]{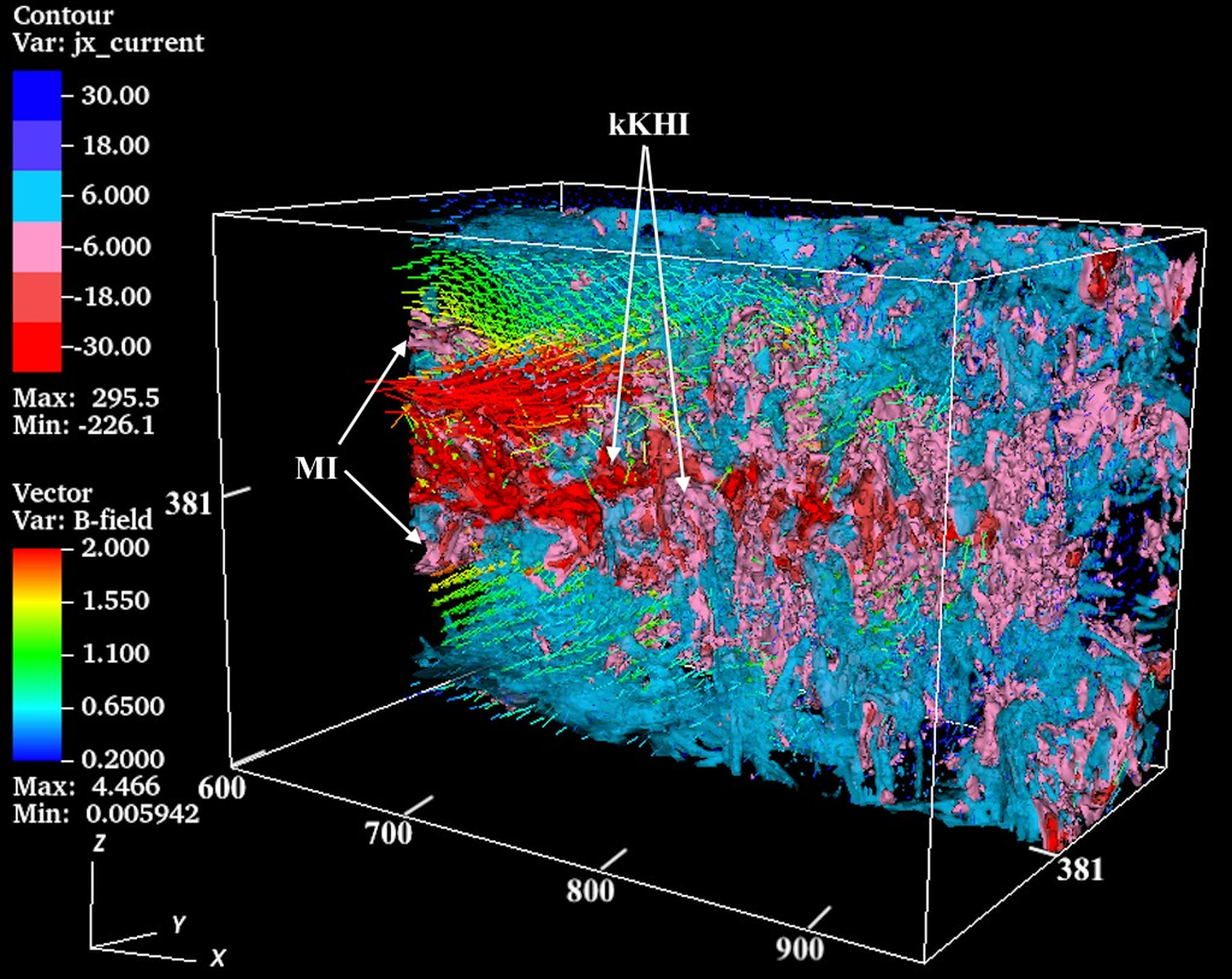}
\end{minipage}

\caption{Current density ($J_x$) overlaid with magnetic field vectors within a
cuboid volume ($600 < x/\Delta < 950$; $256 < y/\Delta,\, z/\Delta < 506$),
showing a cut-out cross-section taken at the jet center
($y/\Delta = z/\Delta = 381$) at $t = 1000\,\omega_{\mathrm{pe}}^{-1}$.
The black boxes in Figs.~\ref{fig:magnetic_evolution_pa}(c, d) and \ref{fig:magnetic_evolution_ra}(c, d) indicate the volume shown in these plots.}
\label{fig:3D-Jx_combined}
\end{figure*}

Figures~\ref{fig:3D_field_evolution_combined}a, b and~\ref{fig:3D_field_evolution_combined}c, d show the three-dimensional
magnetic field structure near the jet head in the region $770 < x/\Delta < 1120$ (indicated by red
boxes in Figs.~\ref{fig:magnetic_evolution_pa}c, d and
\ref{fig:magnetic_evolution_ra}c, d) for $e^{\pm}$ and $e^{-}$ -- $i^{+}$ jets at $t = 1000\omega_{\mathrm{pe}}^{-1}$, respectively.
We observe the  jet head region near $x/\Delta \simeq 1100$ at the  time  $t = 1000\omega_{\mathrm{pe}}^{-1}$ which is consistent with total physical time elapsed and distance travelled by the jet particles. In this region, weak magnetic fields are generated due to kinetic instabilities by jet particles. The particle phase-space distributions shown in Figs.~\ref{fig:Ex_phasepa_combined_all}c ,d and ~\ref{fig:Ex_phas_ra_combined_comparison}c, d provide evidence that jet particles propagate to the vicinity of the jet-head region, reaching approximately $x/\Delta \simeq 1100$. For the $e^{\pm}$ jets,
Figs.~\ref{fig:3D_field_evolution_combined}a and ~\ref{fig:3D_field_evolution_combined}b show that the magnetic fields in
the region $900 < x/\Delta < 1000$ first become twisted and then dissipate.
At approximately $x/\Delta = 950$, the magnetic field generated by the outer
MI mode dissipates.
The magnetic fields associated with the inner MI mode weaken, but reappear
after $x/\Delta \approx 1080$.
For $\gamma_{0} = 15$ (see Fig.~\ref{fig:3D_field_evolution_combined}b), the magnetic field vectors spread
more in the radial direction in the early nonlinear stage than in the
lower-$\gamma_{0}$ jet, indicating delayed magnetic field dissipation for the
higher-$\gamma_{0}$ jet.

For the $e^{-}$ -- $i^{+}$ case, Figs.~\ref{fig:3D_field_evolution_combined}c and ~\ref{fig:3D_field_evolution_combined}d show a weakened magnetic field at
the jet center surrounded by swirling magnetic fields.
The front edge of the toroidal magnetic field at the center is peeled off during
jet propagation, while traces of strong magnetic field remain visible in the
core regions.
This indicates that the toroidal magnetic field may have dissipated in the
nonlinear stage and consequently migrated toward the jet boundary.
We note that the acceleration of jet particles in the nonlinear stage in the
region $900 < x/\Delta < 1100$ is not only due to the electric fields generated
by the instabilities, but is also correlated with the dissipation of magnetic
field.
In the nonlinear stage, the magnetic field generated by kinetic instabilities
dissipates.
Figures~\ref{fig:Ex_phasepa_combined_all}c and ~\ref{fig:Ex_phasepa_combined_all}d show the
acceleration of particles in multiple bunches during the dissipation of the
magnetic field near $x/\Delta = 900$.
The magnetic field dissipation is notably less pronounced in the $e^{-}$ -- $i^{+}$ jet in comparison with the $e^{\pm}$ jets.
 
\begin{figure*}[t!]
\centering
    \makebox[\linewidth]{%
        \makebox[0.46\linewidth]{\phantom{\textbf{(a)}\hspace{1.0cm}}\hfill$\gamma_{0} = 10$\hfill\textbf{(a)}\hspace{1.0cm}}%
        \makebox[0.09\linewidth][c]{\textbf{$e^{\pm}$ jet}}%
        \makebox[0.42\linewidth]{\phantom{\textbf{(b)}\hspace{1.0cm}}\hfill$\gamma_{0} = 15$\hfill\textbf{(b)}\hspace{1.0cm}}%
    }\\[0.05cm]
\begin{minipage}{0.48\textwidth}
  \includegraphics[width=\textwidth]{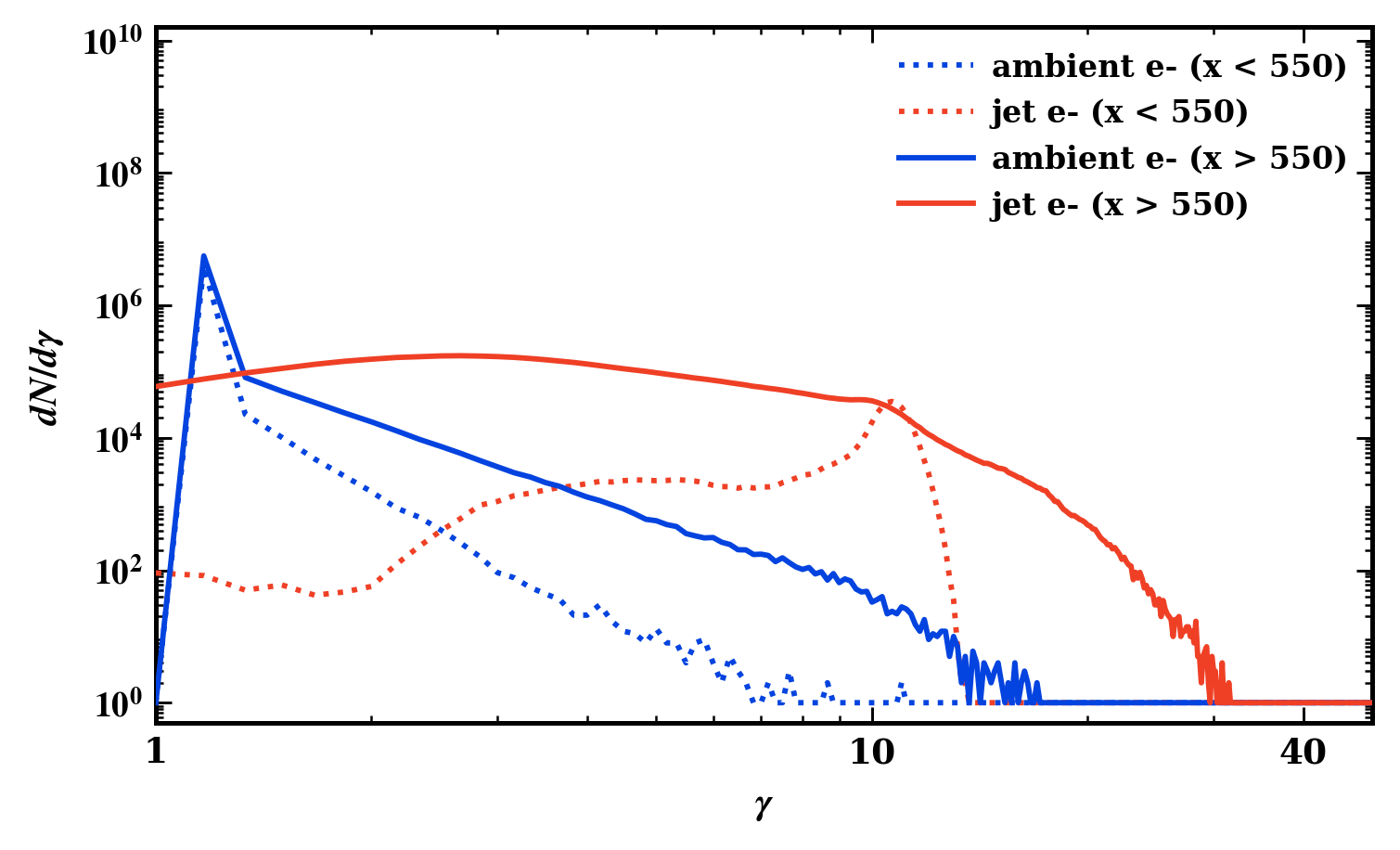}
\end{minipage}
\hfill
\begin{minipage}{0.48\textwidth}
  \vspace{-0.06cm}
  \includegraphics[width=\textwidth]{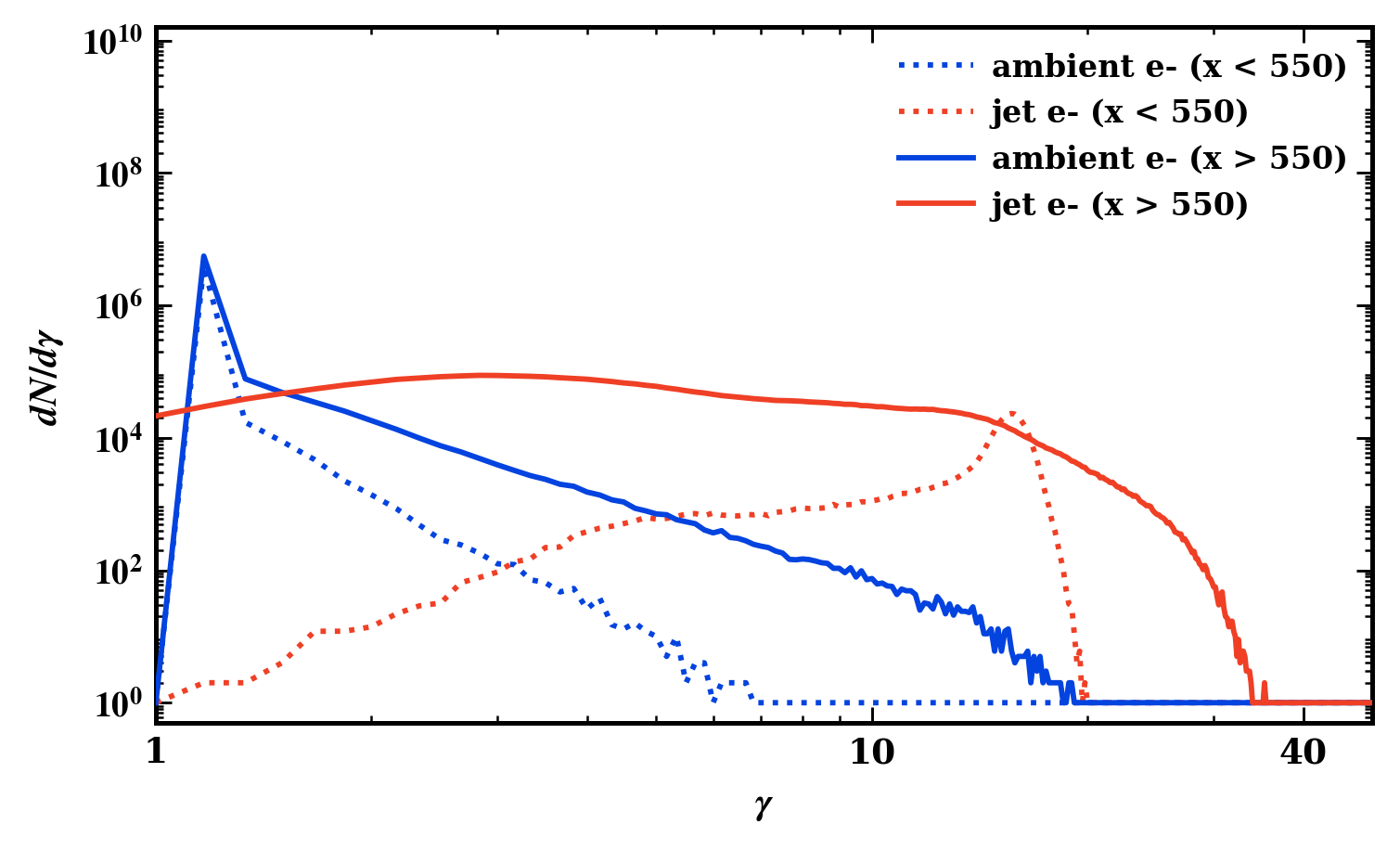}
\end{minipage}

\vspace{-0.06cm}

    \makebox[\linewidth]{%
        \makebox[0.46\linewidth]{\phantom{\textbf{(c)}\hspace{1.0cm}}\hfill\textbf{(c)}\hspace{1.0cm}}%
        \makebox[0.12\linewidth][c]{\textbf{$e^{-}$ -- $i^{+}$ jet}}%
        \makebox[0.42\linewidth]{\phantom{\textbf{(d)}\hspace{1.0cm}}\hfill\textbf{(d)}\hspace{1.0cm}}%
    }\\[-0.06cm]
\begin{minipage}{0.48\textwidth}
  \includegraphics[width=\textwidth]{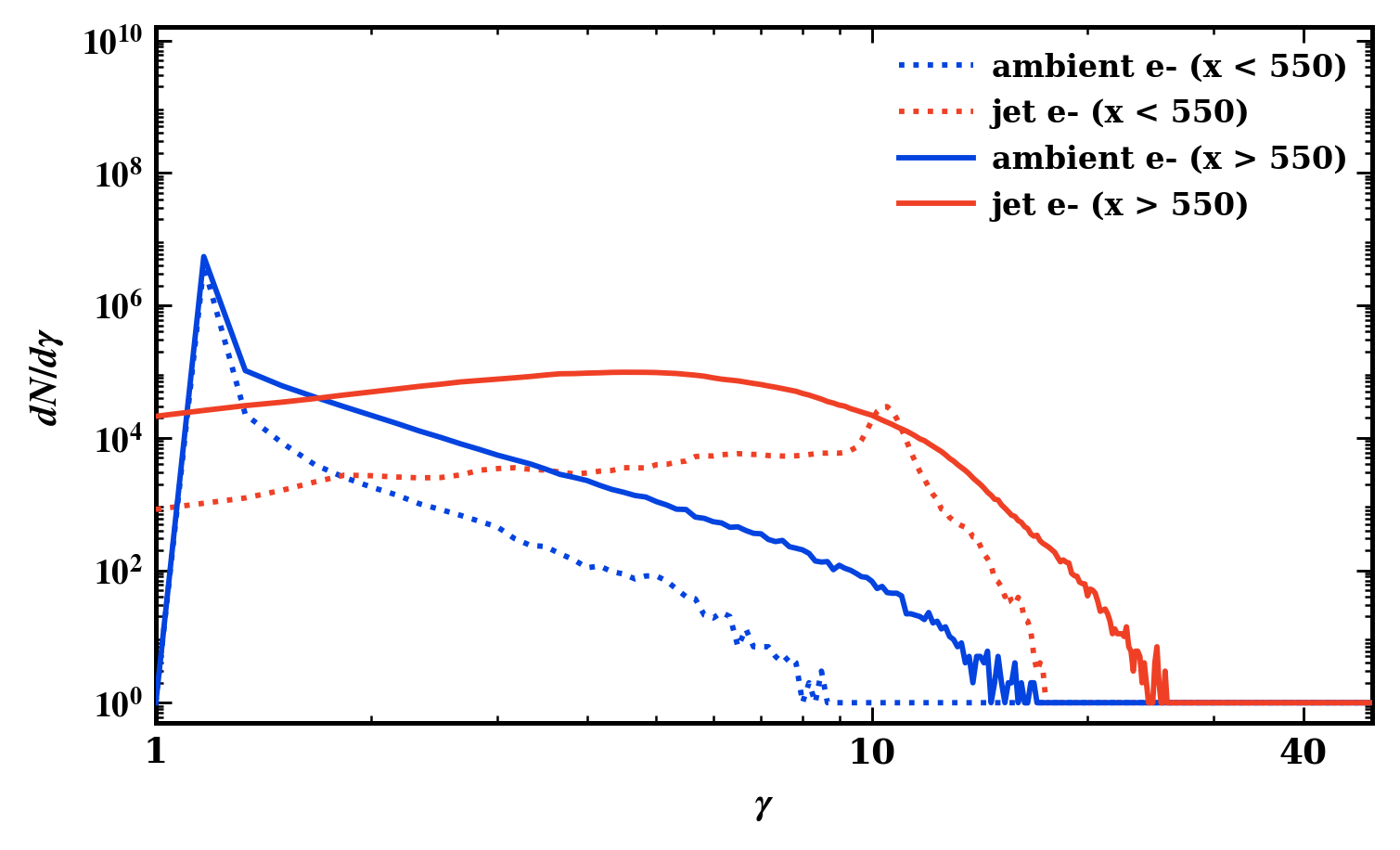}
\end{minipage}
\hfill
\begin{minipage}{0.48\textwidth}
  \vspace{-0.06cm}
  \includegraphics[width=\textwidth]{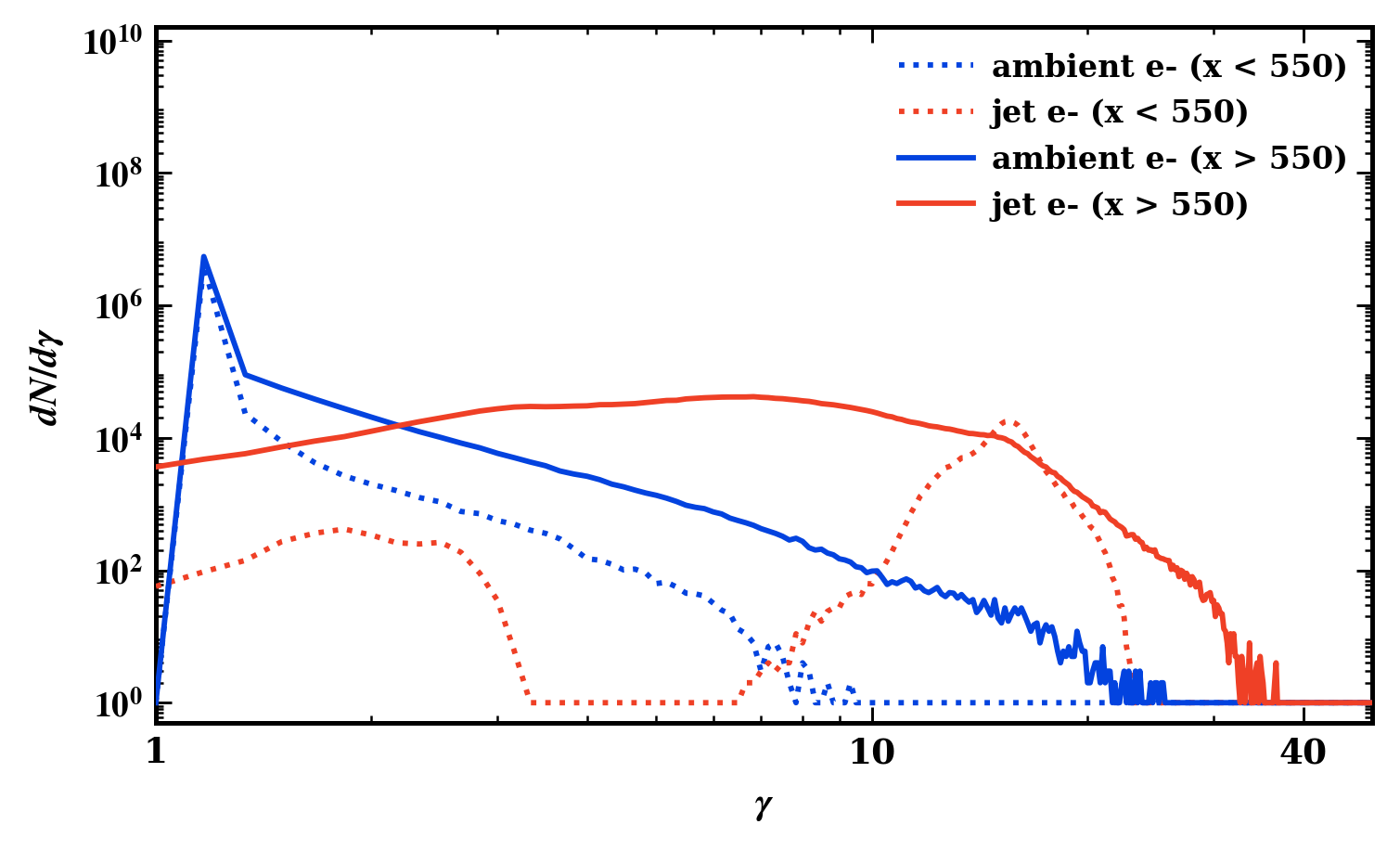}
\end{minipage}

\vspace{-0.06cm}

\caption{Particle energy distributions for jet (red) and ambient plasma (blue)
at $t = 1000\,\omega_{\mathrm{pe}}^{-1}$. In all panels, the spectra are integrated over two distinct spatial
regions: $x/\Delta < 550$ (dashed lines) and $x/\Delta > 550$ (solid lines).}
\label{fig:combined_energy_dist}
\end{figure*}
 
\subsection{Kinetic Instabilities and Particle Energy Distribution}

Figure~\ref{fig:3D-Jx_combined} shows the growth of the kKHI and the
generation of two MI modes in the region $600 < x/\Delta < 950$, extending
along the jet radius ($z$-axis) up to $x/\Delta = 750$ (indicated by black
boxes in Figs.~\ref{fig:magnetic_evolution_pa}c, d and
\ref{fig:magnetic_evolution_ra}c, d at $t = 1000\,\omega_{\mathrm{pe}}^{-1}$.
The corresponding acceleration and deceleration of particles are shown in Figs.~\ref{fig:Ex_phasepa_combined_all}c, d
and \ref{fig:Ex_phas_ra_combined_comparison}c, d.
In both cases, the MI intensifies; consequently, the jet electrons are strongly
collimated.
For the higher-$\gamma_{0}$ jet, around $x/\Delta = 850$, the collimated jet
electron structure weakens and becomes stratified, as shown in 
Figs.~\ref{fig:density_evolution_all}c, ~\ref{fig:density_evolution_all}d.
This indicates the presence of the nonlinear stage of the MI and kKHI.  
Moreover, it clearly illustrates the spatial evolution
	of the saturation and dissipation of kinetic instabilities for different
	plasma compositions and initial Lorentz factors. Compared with Fig.~\ref{fig:3D_field_evolution_combined},
	it reveals more complex turbulent structures in the dissipation region
	ahead of the cone-shaped organized saturated magnetic field. The MI mode near the jet boundary first dissipates at approximately
$x/\Delta = 850$, while the inner mode (MI and kKHI) persists longer and
dissipates at approximately $x/\Delta = 900$, as is more clearly visible in
Fig.~\ref{fig:3D_field_evolution_combined}.

In the nonlinear stage, the MI mode in the jet starts to split, resulting in
several magnetic structures from the remnants of the MI modes, which are seen in Fig.~\ref{fig:3D-Jx_combined}.
These structures are primarily found in regions of weak magnetic fields,
represented by the blue regions in Fig.~\ref{fig:Btotyzisland_combined_2by2}.
In these regions, swirling and oppositely directed magnetic fields exist,
creating a more turbulent magnetic environment.
Figure~\ref{fig:Btotyzisland_combined_2by2} illustrates the typical morphology
of the turbulent magnetic field in all four cases.
Our simulation results demonstrate the formation of magnetic islands with
oppositely swirling magnetic field patterns at regions of minimum field strength
(indicated by dark blue), suggestive of reconnection-like magnetic
restructuring.
These magnetic fields are produced by the jet current, which is modulated by
the kKHI and the MI.
As the outer MI mode dissipates, the magnetic field weakens near the jet
boundary, causing the magnetic field structure to become distorted and
dispersed.
For the $e^{-}$ -- $i^{+}$ jet with $\gamma_{0} = 15$, the central island in
Fig.~\ref{fig:Btotyzisland_combined_2by2} is more spatially
localized and reaches a higher peak field strength ($1.908$) compared with the
$\gamma_{0} = 10$.
This suggests that the initial Lorentz factor directly affects the compression and
stability of the resulting magnetic islands.

Figures~\ref{fig:combined_energy_dist}a, b and~\ref{fig:combined_energy_dist}c, d show the energy distributions of jet
(red) and ambient (blue) electrons in the $e^{\pm}$ jet 
and the $e^{-}$ -- $i^{+}$ jet for the two spatial regions
$x/\Delta < 550$ and $x/\Delta > 550$ at $t = 1000\,\omega_{\mathrm{pe}}^{-1}$, respectively.
We find that electron acceleration is most significant in the nonlinear stage
($x/\Delta > 550$).
Jet electrons reach Lorentz factors of $\gamma = 20$--$38$ for both plasma
compositions.
The jet electrons are mostly accelerated in the nonlinear stage, reaching a
maximum Lorentz factor of $\gamma = 30$--$38$.
For the higher-$\gamma_{0}$ jets ($\gamma_{0} = 15$), the electron acceleration
for both plasma compositions reaches Lorentz factors of $\gamma = 35$--$40$.
However, for the lower-$\gamma_{0}$ case, the jet electron acceleration in the
linear stage reaches only $\gamma \approx 20$.
This indicates that jet electron acceleration in the linear stage is much lower
than in the higher-$\gamma_{0}$ case, consistent with the structural differences
observed in Figs.~\ref{fig:lorentz_electrons_combined} and
\ref{fig:density_evolution_all}.
Comparing the second row of Figs.~\ref{fig:magnetic_evolution_pa} and
\ref{fig:magnetic_evolution_ra} with the bottom panels of
Figs.~\ref{fig:Ex_phasepa_combined_all} and
\ref{fig:Ex_phas_ra_combined_comparison}, we find that the jet electrons are
further accelerated once magnetic field dissipation occurs in the nonlinear
stage, as also observed in kinetic simulations of driven magnetized turbulence
\citep{Meli2023}.

\section{Summary and Discussion}
\label{sec:summary}

We carried out fully self-consistent, three-dimensional relativistic
particle-in-cell (PIC) simulations of both $e^{\pm}$ and $e^{-}$ -- $i^{+}$
plasma jets propagating into an unmagnetized ambient plasma.
The primary objective of this work was to investigate how variations in
the initial Lorentz factor ($\gamma_{0}$) and plasma composition
affected the overall jet morphology and evolution.
Specifically, we examined the excitation of kinetic instabilities and the
resulting particle acceleration driven by the evolving electromagnetic
field topology, including the evolution of the associated electric and
magnetic fields through both linear and nonlinear stages. 

In collisionless relativistic plasmas, shock and shear-driven kinetic instabilities emerge as key mechanisms for magnetic field generation \citep{Nishikawa2016a, Nishikawa2016b, Meli2023}. These processes are considered promising candidates for amplifying magnetic fields in high-energy astrophysical environments, including gamma-ray bursts, relativistic jets from active galactic nuclei, and galaxy clusters. In the present study, the spatial variation of the locally cell-averaged Lorentz factor reveals that
lower initial Lorentz factors ($\gamma_{0} = 10$) promote more pronounced
transverse particle diffusion across both plasma compositions.
This spatial broadening is a combined consequence of weaker relativistic beaming
and enhanced nonlinear growth of shear-driven modes, such as the mushroom
instability. The excited MI and kKHI radially push out jet electrons (see Fig.~\ref{fig:lorentz_electrons_combined}). Because jet ions are heavier, they are not pushed out radially as easily as jet electrons and positrons (see Figs.~\ref{fig:lorentz_positrons_ions_combined}b and ~\ref{fig:lorentz_positrons_ions_combined}d).
Conversely, higher initial Lorentz factors ($\gamma_{0} = 15$) yield
substantially larger maximum attained Lorentz factors, a trend that is
fundamentally modulated by the plasma composition.
We find that the growth of plasma-shear instabilities depends on both the initial Lorentz
factor and the plasma composition of the jet \citep[e.g.,][]{Nishikawa2014}.
For the $e^{\pm}$ jet, in the nonlinear stages, we observe weaker total magnetic field (see Fig.~\ref{fig:magnetic_evolution_pa}e) with
stronger growth of kinetic instabilities (see Fig.~\ref{fig:magnetic_evolution_pa}c) in the lower-$\gamma_{0}$, consistent with the vigorous growth of
the MI \citep[e.g.,][]{Alves2015, Guo2026}.
We observe different behavior for the $e^{-}$ -- $i^{+}$ case: the kinetic
instabilities (see Fig.~\ref{fig:magnetic_evolution_ra}d) in the nonlinear stage are
stronger for the higher-$\gamma_{0}$ jet.
These results show that $\gamma_{0}$ has a significant influence on the $e^{-}$ -- $i^{+}$ jet at a mass ratio of $m_{i}/m_{e} = 4$. The electron density color maps for the higher-$\gamma_{0}$ jet show that the collimation of jet electrons begins earlier, and the strongest collimation is observed in the $e^{-}$ -- $i^{+}$ jet with
$\gamma_{0} = 15$ (see Fig.~\ref{fig:density_evolution_all}d).
We therefore conclude that jet-electron collimation is enhanced in the $e^{-}$ -- $i^{+}$ jet relative to the $e^{\pm}$ case at the same $\gamma_{0}$.
The total magnetic field at $x/\Delta = 1000$ in all four cases reveals a
highly turbulent structure in the late nonlinear stage, near which multiple
acceleration peaks of jet electrons are observed.
These structures support the finding that, after the magnetic field dissipates,
it re-emerges in a reorganized turbulent state with multiple magnetic field
islands.
This is evident in Figs.~\ref{fig:3D_field_evolution_combined}c and~\ref{fig:3D_field_evolution_combined}d, which show that magnetic field dissipation is weaker in the
 $e^{-}$ -- $i^{+}$ jet.
This indicates that, in the nonlinear stage, particle acceleration is not only
due to the growth of kinetic instabilities; the turbulent magnetic field with
multiple magnetic islands also plays an important role.

Figures~ \ref{fig:dEkin_em_comparison}a, b clearly show the trend of conversion of kinetic energy into electromagnetic energy. In the linear stage, the electromagnetic energy is larger at $\gamma_{0} = 10$ than at $\gamma_{0} = 15$ for both the $e^{\pm}$ jet 
	and the $e^{-}$ -- $i^{+}$ jet, providing evidence for larger growth rates of kinetic instabilities at lower initial Lorentz factor. In both the linear
and nonlinear stages, only a few ambient electrons show very slow acceleration. The ambient electrons have broader,
inclined acceleration peaks that trail the jet front due to their initial low velocities. They appear prominently in the dissipation region, and halt earlier than the jet electrons. Positron acceleration and deceleration remain significant even for slightly smaller electric fields because positrons are much lighter than ions.
The particle energy distribution curves show that
acceleration is most significant in the nonlinear stage ($x/\Delta > 550$; see Fig.~\ref{fig:combined_energy_dist}).
We observe significant acceleration for $\gamma_{0} = 15$ in both jet compositions, reaching Lorentz
factors of $\gamma \approx 35$--$38$ as in Figs.~\ref{fig:combined_energy_dist}b and~\ref{fig:combined_energy_dist}d. The self-consistently generated turbulent magnetic field with
multiple magnetic field islands, as shown in
Fig.~\ref{fig:Btotyzisland_combined_2by2}, supports turbulent particle
acceleration in the nonlinear stage. 

Whether AGN jets are pair-dominated or baryon-loaded remains an important open question for studies of plasma kinetic instabilities, jet collimation, and high-energy emission \citep[e.g.,][]{Alves2018, Alves2019,Ortuo-Macas2022, Meli2023}. Polarization diagnostics using RMHD simulations can discriminate the two cases \citep{Anantua2020}, while kinetic studies suggest that composition may evolve with radius through pair enrichment \citep{Mehlhaff2025}. Future work will combine these kinetic results with radiative-transfer and polarimetric modeling to enable direct comparison with multiwavelength observations from facilities such as IXPE, ALMA, HHT and ngEHT \citep[e.g.,][]{Lu2023, Marshall2024, Agudo2025, Goddi2025}.

The present work is a first step toward simulation studies in both
the linear and nonlinear stages with different plasma compositions, which will
address different modes of kinetic instabilities across both regimes.
Although our current simulations are limited by box size, simulation time, and
fixed toroidal field strength ($B_0 = 0.5$ in normalized code units), they offer important insights into
the growth of kinetic instabilities, the generation of electric and magnetic
fields, and particle acceleration in jets with different initial Lorentz factors
and plasma compositions.

We plan to run further simulations with a larger jet radius for sufficiently
long durations.
These will further illuminate the evolution of astrophysical jets, the
dissipation of toroidal magnetic fields, the re-emergence of turbulent magnetic
fields, possible reconnection-like events, and particle acceleration, which are
critical to observational astronomy.
In our forthcoming simulations we will use ion-to-electron mass ratios of
$m_{i}/m_{e} = 16$, $25$, etc., and varied magnetic field strengths, with the aim
of assessing how jet dynamics and instability evolution depend on plasma
composition.
These simulations will allow us to characterize differences in jet structure,
magnetic reconnection sites, and particle acceleration mechanisms under a range
of physical conditions relevant to the observational signatures of astrophysical
jets.

\begin{acknowledgments}
R.C.P., A.M., and K.N. acknowledge the support of the National Science
Foundation (NSF 20-542) under the RISE Foundation.

I.D. acknowledges support from the Romanian Ministry of Research, Innovation and Digitalization under the Romanian National Core Program LAPLAS VII $-$ contract no. 30N$/$2023.

C.K. is funded by the Independent Research Fund Denmark (grant 1054-00104) and the Deutsche Forschungsgemeinschaft (DFG, German Research Foundation),
grant 548340212.
\end{acknowledgments}

\section*{Author ORCID IDs}
Ram Chandra Pageni: \url{https://orcid.org/0009-0002-6873-1259} \\
Kenichi Nishikawa: \url{https://orcid.org/0000-0001-6031-7040} \\
Athina Meli: \url{https://orcid.org/0000-0002-3064-5307} \\
Ioana Duțan: \url{https://orcid.org/0000-0003-3932-5205} \\
Jacob Trier Frederiksen: \url{https://orcid.org/0000-0002-3560-0044} \\
Christoph Köhn: \url{https://orcid.org/0000-0002-7101-5889} \\
Nicholas R. MacDonald: \url{https://orcid.org/0000-0002-6684-8691} \\
Yosuke Mizuno: \url{https://orcid.org/0000-0002-8131-6730} \\

\section{Software and Data Availability}
The simulations were performed with \texttt{TRISTAN-MPI} \citep{Nishikawa2009, Nishikawa2021} on the Frontera supercomputer at the Texas Advanced Computing Center (TACC). Data analysis and visualization used NCAR Graphics \citep{NCAR2024}, VisIt \citep{Childs2012}, and Python with Matplotlib \citep{Hunter2007}, pandas \citep{mckinney2010data}, and NumPy \citep{harris2020array}.

\bibliographystyle{aasjournal}
\bibliography{1_Impact_sub_revR2}
\end{document}